\documentclass[fleqn,usenatbib]{mnras}

\usepackage{newtxtext,newtxmath}

\usepackage[T1]{fontenc}
\usepackage{ae,aecompl}

\usepackage{graphicx}	
\usepackage{amsmath}	
\usepackage{amssymb}	
\usepackage[]{units}
\usepackage{times}

\usepackage{etoolbox}
\makeatletter
\patchcmd\@combinedblfloats{\box\@outputbox}{\unvbox\@outputbox}{}{\errmessage{\noexpand\@combinedblfloats could not be patched}}
\makeatother

\newcommand{\mspc}{M_\odot\,\mathrm{pc}^{-2}}
\newcommand{\dndm}{\frac{\mathrm{d}N_{\rm cl}}{\mathrm{d}M_{\rm cl}}}
\newcommand{\FIREurl}{\url{http://fire.northwestern.edu}}
\newcommand{\gizmourl}{\url{www.tapir.caltech.edu/~phopkins/Site/GIZMO.html}}

\defcitealias{kruijssen:2012.cluster.formation.efficiency}{K12}

\title[Formation of stellar  associations and clusters from GMCs]{A model for the formation of stellar associations and clusters from giant molecular clouds}

\author[M. Y. Grudi\'{c} et al.]{Michael Y. Grudi\'{c}$^{1}$\thanks{E-mail: mike.grudic@northwestern.edu},
J.~M.~Diederik Kruijssen$^{2}$,
Claude-Andr\'{e} Faucher-Gigu\`{e}re$^{1}$,\newauthor
Philip F. Hopkins$^{3}$,
Xiangcheng Ma$^{4}$,
Eliot Quataert$^{4}$, and Michael Boylan-Kolchin$^{5}$
\\
$^{1}$Department of Physics and Astronomy and CIERA, Northwestern University, 2145 Sheridan Road, Evanston, IL 60208, USA\\
$^{2}$Astronomisches Rechen-Institut, Zentrum f\"{u}r Astronomie der Universit\"{a}t Heidelberg, M\"{o}nchhofstra\ss e 12-14, 69120 Heidelberg, Germany\\
$^{3}$TAPIR, Mailcode 350-17, California Institute of Technology, Pasadena, CA 91125, USA\\
$^{4}${Department of Astronomy and Theoretical Astrophysics Center, University of California Berkeley, Berkeley, CA 94720}\\
$^{5}$Department of Astronomy, The University of Texas at Austin, 2515 Speedway, Stop C1400, Austin, TX 78712, USA \\
}

\date{Accepted XXX. Received YYY; in original form ZZZ}

\pubyear{2020}
\begin{document}
\label{firstpage}
\pagerange{\pageref{firstpage}--\pageref{lastpage}}
\maketitle

\begin{abstract}
We present a large suite of MHD simulations of turbulent, star-forming giant molecular clouds (GMCs) with stellar feedback, extending previous work by simulating 10 different random realizations for each point in the parameter space of cloud mass and size. It is found that once the clouds disperse due to stellar feedback, both self-gravitating star clusters and unbound stars generally remain, which arise from the same underlying continuum of substructured stellar density, ie. the hierarchical cluster formation scenario. The fraction of stars that are born within gravitationally-bound star clusters is related to the overall cloud star formation efficiency set by stellar feedback, but has significant scatter due to stochastic variations in the small-scale details of the star-forming gas flow. We use our numerical results to calibrate a model for mapping the bulk properties (mass, size, and metallicity) of self-gravitating GMCs onto the star cluster populations they form, expressed statistically in terms of cloud-level distributions. Synthesizing cluster catalogues from an observed GMC catalogue in M83, we find that this model predicts initial star cluster masses and sizes that are in good agreement with observations, using only standard IMF and stellar evolution models as inputs for feedback. Within our model, the ratio of the strength of gravity to stellar feedback is the key parameter setting the masses of star clusters, and of the various feedback channels direct stellar radiation (photon momentum and photoionization) is the most important on GMC scales.
\end{abstract}

\begin{keywords}
galaxies: star formation -- galaxies: star clusters: general -- stars: formation
\end{keywords}



\section{Introduction}
Star formation (SF) is a statistically-correlated process. It is in some sense \textit{clustered}, with most stars forming as part of stellar structures of elevated stellar density  \citep{lada:2003.embedded.cluster.review,mckee:2007.review,krumholz:2018.star.cluster.review}. This clustering in space is accompanied by clustering in time, with the age spread of a population monotonically increasing with the size of the region over which it is measured \citep{efremov.elmegreen:1998}. In local galaxies, SF is dominated by massive complexes formed in the most massive giant molecular clouds (GMCs), of mass scale $10^6-10^7 M_\odot$ in local galaxies \citep{bolatto:2008.gmc.properties, paws,miville:2017.gmcs,freeman:2017.m83.gmcs}. 
Many important questions remain regarding the detailed relationship between giant molecular clouds, young stellar associations, and the subset of stars formed that are in gravitationally-bound star clusters.

The most basic question is what fraction of the gas mass of of a GMC is converted to stars, the {\it star formation efficiency} (SFE). Observationally, this can be inferred from measurements of the present gas mass and stellar mass in star-forming regions. Although such estimates of the ratio of young stellar to gas mass have yielded instantaneous SFEs ranging over many orders of magnitude ($\sim 10^{-4}-1$) \citep[e.g.][]{myers:1986.gmcs,lee:2016.gmc.eff,vuti:2016.gmcs}, observations appear to be consistent with a picture where GMCs convert a few per cent of their mass into stars, once the scatter due to molecular cloud evolution is accounted for (see \citealt{feldmann:2011}, \citealt{lee:2016.gmc.eff}, \citealt{grudic:2018.gmc.sfe}, \citealt{kruijssen:2019.ngc300}, \citealt{chevance:2020.gmcs}). This efficiency is largely consistent with theoretical models wherein GMCs are disrupted by feedback from main sequence massive stars \citep{williams.mckee:1997, krumholz.matzner.mckee:2006, fall:2010.sf.eff.vs.surfacedensity, murray:molcloud.disrupt.by.rad.pressure, kim:2018,chevance:2020.gmc.review}. In a previous study \citep{grudic:2016.sfe}, we presented MHD simulations of star formation in gas clouds with a wide range of parameters including multiple feedback channels, finding that the cloud surface density sets the relative strengths of feedback and gravity, and hence the SFE achieved before the cloud is disrupted. In \citet{grudic:2018.gmc.sfe}, we found that this picture successfully reproduces the measured instantaneous star formation efficiencies of GMCs in the Milky Way in detail (see also \citealt{geen:2017}).

A complete theory of star formation in GMCs must go further, describing not only how many stars form, but how those stars are organized spatially and dynamically. Regarding this, an important distinction to be made is between {\it monolithic} and {\it hierarchical} pictures of star formation. In the monolithic scenario, all stars originate in star clusters of high stellar density, most of which subsequently expand and dissolve into the field due to gas expulsion and N-body evolution \citep{lada:1991, kroupa:2001.orion.pleiades}. In the hierarchical scenario, stellar structure is inherited from a hierarchy of ISM structure spanning a wide range of scales. As a result, stars form with a wide range of natal stellar densities, and gravitationally-bound, dynamically-relaxed clusters are simply the result of the upper tail-end of a larger continuum that happens to have locally-high star formation efficiency  \citep{elmegreen:2002,elmegreen:2008, bonnell:2011,kruijssen:2012.cluster.formation.efficiency}. The observational case for the latter scenario has strengthened in recent years, with the general finding that young stellar structure does indeed appear to span a broad range of scales and a continuum of densities \citep{bastian:2005.m51.clusters, bressert:2010.clustered.sf, gouliermis:2015.hierarchical.clustering,grasha:2017.hierarchical.clustering,gouliermis:2018.review}. Recently it was also found that the kinematics of OB assocations are incompatible with the hypothesis that they consist only of formerly-dense clusters that have since expanded \citep{ward.kruijssen:2018,ward.kruijssen:2019}. 

The next important question after the SFE is the fraction of the stars locked into bound clusters at the end of star formation:
\begin{equation}
    f_{\rm bound} =  \frac{M_{\rm \star,bound}}{M_\star}.
\end{equation}
It has long been thought that the value of $f_{\rm bound}$ is typically $\ll1$, at least in the conditions of local galaxies \citep{elmegreen:1983}. Simple physical arguments can be made that $f_{\rm bound}$ is an increasing function of the local star formation efficiency \citep{hills:1980,mathieu:1983}, which have been refined by N-body experiments \citep{fellhauer.kroupa:2005,baumgardt.kroupa:2007,smith:2011.cluster.assembly, smith:2013.cluster.assembly}. In the monolithic picture, $f_{\rm bound}$ corresponds to the fraction of clusters that survive ``infant mortality", the expansion that occurs when gas within the cluster is expelled by stellar feedback. Within the hierarchical picture, it has been argued that bound clusters form in regions where feedback is inefficient, \textit{exhausting} gas locally so that they generally avert infant mortality \citep{kruijssen:2012.gas.exhaustion, dale:2015.winds.ionizing, 2016A&A...595A..27G}. Meanwhile, there would be a population of stars that never get a chance to orbit within a bound, virialized structure in the first place -- in this scenario, $f_{\rm bound}$ corresponds to the mass fraction of the population that {\it does} exist within a bound cluster \citep{bastian:2008.cfe, kruijssen:2012.cluster.formation.efficiency}.
 
It has historically been quite difficult to reliably measure $f_{\rm bound}$ through direct observations of any one star-forming cloud complex, as very good astrometric data are needed\footnote{However, recently progress has been made in estimating the boundedness of young star clusters in the Milky Way, see e.g.\ \citet{ginsburg:2018.cfe} and \citet{kuhn:2018.gaia}.}. On the other hand, under certain assumptions, it is possible to measure an average $f_{\rm bound}$ integrated over an entire galaxy or a patch of a galaxy \citep{bastian:2008.cfe, goddard:2010.cfe}:
\begin{equation}
   \Gamma = \frac{\dot{M}_{\rm bound}}{\dot{M}_{\star}},
\end{equation}
where $\dot{M}_{\rm bound}$ and $\dot{M}_{\star}$ are the mass formation rates of stars in bound clusters and of all stars in the region. Extragalactic studies can only determine the formation rate of stars in dense stellar structures, $\dot{M}_\mathrm{dense}$, by measuring the mass in dense stellar structures within a certain age bin, and stellar density is not necessarily a sufficient condition for boundedness. However, it is a reasonable approximation if measured over a proper choice of age bin, using only clusters that are too old to have survived as an unbound entity (i.e.\ older than their internal crossing time, $\sim \unit[1]{Myr}$), but too young to be likely to be disrupted or to have lost much mass due to galactic or internal dynamical processes ($\leq \unit[100]{Myr}$) \citep[e.g.][]{kruijssen.bastian:2016.pitfalls}. This measurement has been performed most convincingly over different regions of several local spiral galaxies \citep{adamo:2015.m83.clusters, johnson:2016.cluster.formation.efficiency, messa:2018.m51}, with the general finding that regions of greater mean ISM pressure, molecular gas fraction, and gas surface density tend to have greater values of $\Gamma$, in line with analytic expectations invoking the progressive inefficiency of stellar feedback toward higher densities \citep{kruijssen:2012.cluster.formation.efficiency}.

Of the material that does remain bound, one must then ask what the properties of the remaining bound structures are. The maximum bound cluster mass in particular is expected to be a sensitive probe of star formation physics, because if GMCs constitute the gas supply potentially available for cluster formation, bound cluster mass satisfies \citep{kruijssen:2014.cluster.formation}:
\begin{equation}
    M_{\rm cl} \leq f_\mathrm{bound} \epsilon_{\rm int} M_{\rm GMC,max},
    \label{eq:mclmax}
\end{equation}
where $M_{\rm GMC,max}$ is the maximum GMC mass in a given environment, and
\begin{equation}
    \epsilon_{\rm int} = \frac{M_\star}{M_{\rm GMC}}    
    \label{eq:epsint_def}
\end{equation}
is the integrated star formation efficiency of the cloud. Both $f_\mathrm{bound}$ and $\epsilon_\mathrm{int}$ are expected to be sensitive to the strength of stellar feedback, so $M_\mathrm{cl}$ is {\it doubly} sensitive. Note that Equation \ref{eq:mclmax} does not necessarily hold as an equality because one GMC can potentially produce multiple bound clusters -- GMCs can and do produce star cluster \textit{complexes}, where the distribution of cluster masses is described by some underlying distribution.

In this paper we use numerical simulations to approach the above questions about the nature of star cluster formation. We extend our previous work that focused on various aspects of the SFE of GMCs \citep{grudic:2016.sfe,grudic:2018.gmc.sfe, hopkins.grudic:2018.rp, elephant}, using an expanded suite of numerical simulations to map out the behaviour of star-forming GMCs across parameter space, and, crucially, across 10 different random realizations for the initial turbulent flow of each cloud model. From these numerical results we construct a model that predicts the following statistical properties of star cluster populations formed in GMCs:
\begin{itemize}
    \item Their star formation efficiency.
    \item The fraction of stars formed that are locked into gravitationally-bound clusters.
    \item The mass function of bound star clusters, determined at the level of individual clouds.
    \item The size-mass relation of bound clusters.
    \item The initial density profiles of bound clusters.
\end{itemize}
In doing so, we link the physics of MHD turbulence, gravity, radiative processes, star formation, and stellar feedback to the observables that provide the most sensitive probes of star formation physics, in a self-consistent framework, and show that this framework contains the necessary and sufficient ingredients to reproduce these key observations.

This paper is structured as follows. In \S\ref{section:sims}, we describe our simulation and analysis methods, and present the raw numerical results of the study. In \S\ref{section:model}, we describe an analytic statistical model for mapping clouds onto cluster populations that reproduces the results of the simulations, and can be applied to a general cloud population.  In \S\ref{section:M83}, we compare model predictions with observations in M83, and demonstrate that the model can recover realistic star cluster properties from observed GMC properties. In \S\ref{sec:discussion}, we discuss various predictions and applications of our model, and compare it with other models. Finally, in \S\ref{sec:conclusion}, we summarize our main findings and outline future work.

\section{Simulations} \label{section:sims}
\subsection{Numerical Methods}
We perform a suite of simulations of isolated GMCs with {\small GIZMO}, a mesh-free, Lagrangian finite-volume Godunov code designed to capture the advantages of both grid-based and smoothed-particle hydro-dynamics (SPH) methods, described fully in \citet{hopkins:gizmo} \footnote{A public version of this code is available at \gizmourl.}. We solve the equations of ideal magnetohydrodynamics using the Lagrangian Meshless Finite Mass (MFM) method \citep{hopkins:gizmo.mhd}, augmented with a novel constrained-gradient method to further reduce $\nabla \cdot \mathbf{B}$ errors \citep{hopkins:2016.divb}.

\subsubsection{Gravity}
The gravitational field is summed using the fast, approximate \citet{barneshut} tree algorithm introduced in {\small GADGET-3} \citep{springel:gadget}. However, we have modified the original node-opening criterion used in the {\small GADGET-3} algorithm for our problem. In the original algorithm, a node was opened if the estimated field contribution of its quadrupole moment was greater than some small fraction of the total field at the point of interest. In our problem, this can allow the external tree-force on a dense star cluster to be degraded in accuracy due to its locally-strong gravitational field. To avoid this, we enforced the \citet{barneshut} geometric node-opening criterion with an opening angle $\Theta = 0.5$, in addition to the standard \citet{springel:gadget} criterion. The gravitational softening of both gas and star particles is adaptive, with correction terms to ensure energy and momentum conservation as as described in \citet{hopkins:gizmo}. A minimum Plummer-equivalent softening of $\unit[10^{-2}]{pc}$ is enforced only for star particles, however we found that the stellar densities needed for this to have a significant effect are almost never achieved in our parameter space.  

\subsubsection{Star formation}
Our simulations do not attempt to resolve the formation, motion, and feedback of \textit{individual} stars. Rather, as in \citet{grudic:2016.sfe}, they simulate the stellar mass distribution as an ensemble of collisionless star particles. In \citet{grudic:2017} we found that this simulation technique succeeds at producing star clusters of a similar density profile shape to observed young, massive star clusters in local galaxies. The star formation criteria are as described in \citet{grudic:2016.sfe}: gas cells may only be converted to stars if they are self-gravitating at the resolution scale (the virial criterion, \citealt{hopkins:virial.sf}), molecular, and in a converging flow. 

We do not impose a threshold density for star formation. In our initial experiments varying the density threshold, we found that it was either irrelevant compared to the virial criterion if set low, or clearly imprinted a characteristic, numerically-imposed 3D density on the star clusters if set high. We thus decided to rely mainly on the virial criterion, which is has more physical motivation. However, we initially found that this alone was not strict enough, because momentary noise in the velocity gradient could potentially allow premature star formation, and convergence of cloud star formation histories with resolution was slow because the low-resolution runs would form stars systematically sooner. We therefore adopted a smoothing procedure for the virial criterion. If $\alpha_{\rm vir}=\frac{E_{\rm  kin}}{|E_{\rm  grav}|}$ is the local virial parameter, then at each timestep we update dimensionless quantity
\begin{equation}
    A\left(t+\Delta t\right) = \frac{\Delta t}{\tau} \frac{1}{1+\alpha_{\rm vir}\left(t\right)} + \left(1 - \frac{\Delta t}{\tau}\right) A\left(t\right),
\end{equation}
ie. $A\left(t\right)$ is the quantity $\frac{1}{1+\alpha_{\rm vir}} \in \left[0,1\right]$ exponentially smoothed with an $e$-folding time $\tau$. Star formation is allowed when $A\left(t\right) > \nicefrac{1}{2}$, corresponing to $\alpha_{\rm vir} < 1$. We found that setting $\tau = t_{\rm ff}/8$, i.e. smoothing over a window just $\nicefrac{1}{8}$ the local freefall time, is sufficient to de-noise the virial criterion. This was necessary to obtain star cluster mass functions that are robust to numerical resolution (see Appendix \ref{section:appendix}).

\subsubsection{Cooling and Stellar Feedback}
Our treatment of ISM physics and stellar feedback largely follows the FIRE-2 simulations\footnote{\FIREurl}, and all algorithms are presented in detail in \citet{fire2}. We account for an extensive range of radiative cooling and heating processes, using a standard implicit algorithm, and follow cooling down to a numerically-imposed floor of $\unit[10]{K}$.

We include all important nuclear-powered stellar feedback mechanisms from main-sequence massive stars: stellar winds, radiation, and supernova explosions, all of which are dominated by the most massive ($\geq \unit[8]{M_\odot}$) stars for young stellar populations. Each star particle is assigned feedback fluxes consistent with a simple stellar population with a well-sampled \citet{kroupa:imf} IMF, with luminosities, mass loadings, and momentum fluxes taken from a {\small STARBURST99} \citep{starburst99} stellar population model (see \citet{fire2} for details). 

We caution that the results of GMC simulations with stellar feedback {\it do} depend somewhat on the specific prescription for distributing feedback fluxes among star particles, even assuming a given IMF \citep{elephant}. However, here we will only target the parameter space of relatively massive ($>10^6 M_\odot$) GMCs, where cloud lifetimes are long compared to the formation time of massive stars, and the IMF should indeed be well-sampled. As such, we expect that the results of this paper are much less sensitive to the small-scale details of how massive star formation is modeled than the less-massive ($10^5M_\odot$) cloud simulated in \citet{elephant}.

Mass, energy, and momentum fluxes from stellar winds and supernova explosions are injected into particles within the hydrodynamic stencil surrounding a star particle according to the fully-conservative scheme described in \citet{hopkins:2018.sne}. The energy-conserving Sedov-Taylor phase of individual supernova blast-waves is resolved explicitly in all simulations, with the most coarsely-resolved runs having a mass resolution of $\unit[140]{M_\odot}$, and most others much finer, down to $2\,M_\odot$.

We use the {\small LEBRON} radiative transfer approximation (for details see \citealt{fire2}), which accounts for absorption of single-scattered photons within the a stencil around a star particle (including ionizing radiation, expanding the search radius until ionizing photons are exhausted). Photons not absorbed within the stencil are propagated through the simulation domain using an the optically-thin approximation that uses the gravity solver, with extinction corrections at the source and absorber. The radiation field is computed in far-UV, near-UV, optical/NIR and FIR bins, subject to extinction according to appropriate flux-mean opacities, including dust extinction. We have shown in previous work that this radiative transfer approximation gives cloud SFEs in reasonable agreement (factor of $\sim 2$) with results from an {\small M1} closure scheme (e.g. \citealt{rosdahl:2015.rhd}, as demonstrated in \citealt{hopkins.grudic:2018.rp, hopkins:2020.fire.rt}), and even better agreement with the results of a state-of-the-art adaptive ray tracing scheme (e.g. \citealt{kim:2018}, as shown in \citealt{elephant}).

\subsubsection{Cluster identification}
\label{section:clusterfinding}
We identify self-gravitating star clusters at the end of the simulations with our own version of the {\small SUBFIND} algorithm \citep{springel:2001.subfind}. This algorithm organizes the stellar density field into a hierarchy of structures surrounding density peaks, and then at each level of the hierarchy subjects the structures to an iterative unbinding procedure to determine each group's gravitationally-bound subset. Cluster membership is determined according to the \textit{smallest} structures in this hierarchy that can be constructed. We use fast neighbor-lookup and gravitational potential routines provided by {\small scipy} \citep{scipy} and {\small pykdgrav}\footnote{\url{https://github.com/mikegrudic/pykdgrav}}, respectively. We have also run our analysis using the simpler cluster-finding algorithm described in \citet{grudic:2017} based on grouping stars into common potential wells, and have found that none of the results of this study were senstive to the choice of algorithm.

\subsection{Initial Conditions}
Our initial conditions consist of a spherical cloud of uniform density with total mass $M_{\rm GMC}$ and radius $R_{\rm GMC}$, embedded in a warm, diffuse medium in thermal pressure equilibrium that fills a periodic box of side length $20 R_{\rm GMC}$. The initial velocity field of the gas in the cloud consists entirely of a Gaussian random field with power spectrum $|\mathbf{\tilde{v}}\left(\mathbf{k}\right)|^2 \propto k^{-4}$ \citep{gammie.ostriker:1996.mhd.dissipation}, and a natural mixture of solenoidal and compressive modes (ie. $E_{\rm solenoidal} = 2 E_{\rm compressive}$). The velocity field is normalized so that the initial kinetic energy of the cloud is equal in magnitude to its gravitational potential energy. This is a more realistic model of GMCs than the numerical experiments in \citet{grudic:2018.gmc.sfe}, which were intended to isolate a certain surface density scale by ensuring rotational support, which GMCs do not generally have \citep{braine:2018.gmcs}. The warm, diffuse medium is initially at rest. The magnetic field is initially uniform, and normalized so that the magnetic energy in the cloud is $1\%$ of its turbulent energy.

\begin{table}
    \setlength\tabcolsep{3.0pt} 
	\centering
		\begin{tabular}{|c|c|c|c|c|c|}
		\hline
		$M_\mathrm{GMC}$ ($M_\odot$) & $R_\mathrm{GMC}$ (pc) & $\Sigma_\mathrm{GMC}\,\left(\mspc\right)$ & $\alpha_\mathrm{turb}$ & Metallicities ($Z_\odot$) & Realizations\\
		\hline
	    $2\times 10^6$ & 100 & 63 & 2 & 0.01 \& 1 & 10\\ 
	    $4\times 10^6$ & 100 & 130 & 2 & 0.01 \& 1  & 10 \\ 
	    $8\times 10^6$ & 100 & 250 & 2 & 0.01 \& 1 & 10 \\ 
	    $1.6\times 10^7$ & 100 & 500 & 2 & 0.01 \& 1 & 10\\ 
	    $1.8\times 10^7$ & 300 & 63 & 2 & 0.01 \& 1 & 10\\ 
	    $3.6\times 10^7$ & 300 & 130 & 2 & 0.01 \& 1  & 10 \\ 
	    $7.2\times 10^7$ & 300 & 250 & 2 & 0.01 \& 1 & 10 \\ 
	    $1.4\times 10^8$ & 300 & 500 & 2 & 0.01 \& 1 & 10\\ 
		\hline
		\end{tabular}
        \vspace{-0.1cm}
 \caption{Parameter and statistical space of GMC models simulated in this work: initial GMC mass $M_\mathrm{GMC}$, radius $R_\mathrm{GMC}$, mean surface density $\Sigma_\mathrm{GMC}=M_\mathrm{GMC}/\uppi R_\mathrm{GMC}^2$, turbulent virial parameter $\alpha_\mathrm{turb}=\frac{10 E_\mathrm{turb} R_\mathrm{GMC}}{3 G M_\mathrm{GMC}^2}$, metallicity, and the number of independent simulations at a given point in parameter space with different random seeds for the initial velocity field.}
 \label{table:ics}
\end{table}

Our parameter space consists of clouds on a $4\times 2 \times 2$ grid of surface densities ($\unit[64]{\mspc}$, $\unit[127]{\mspc}$, $\unit[254]{\mspc}$, and $\unit[509]{\mspc}$), radii ($\unit[100]{pc}$ and $\unit[300]{pc}$), and metallicities ($\unit[0.01]{Z_\odot}$ and $Z_\odot$) (see Table \ref{table:ics}). We intentionally targeted a parameter space representative of the largest and most massive GMCs in local spiral galaxies \citep[e.g.][]{paws, rice:2016.gmcs,miville:2017.gmcs, freeman:2017.m83.gmcs}, for two reasons. We expect that the most massive (and hence most detectable) clusters originate in the most massive GMCs, and that the total mass of stars formed is dominated by the most massive star-forming complexes \citep[e.g.][]{williams.mckee:1997,murray:2010.sfe.mw.gmc,lee:2016.gmc.eff}. Hence overall, it is likely that the most massive GMCs produce the dominant contribution to the top end of the observable star cluster mass function, an observation that we will confront in  \S\ref{section:M83}. We also expect that this regime of massive star cluster formation is the regime in which our approximations of collisionless stellar dynamics and IMF-averaged feedback are most suitable. Our choice of metallicities is intended to bracket the the range of metallicities of observed globular clusters and galaxies \citep{sluggs,kirby:2013.metallicity}.

In all simulations, the initial cloud component is resolved in $10^6$ Lagrangian gas cells of equal mass. For each point in parameter space, we simulate 10 different random realizations of the initial turbulent velocity field. In doing so, we map out not only the \textit{scalings} across parameter space, but the range of intrinsic cloud-to-cloud variations due to small-scale details of the turbulent gas flow, which can potentially be important for star cluster formation.

\section{Simulation Results}
\subsection{Global Evolution}
\label{section:sims:results}
\begin{figure*}
\centering
\includegraphics[width=\textwidth]{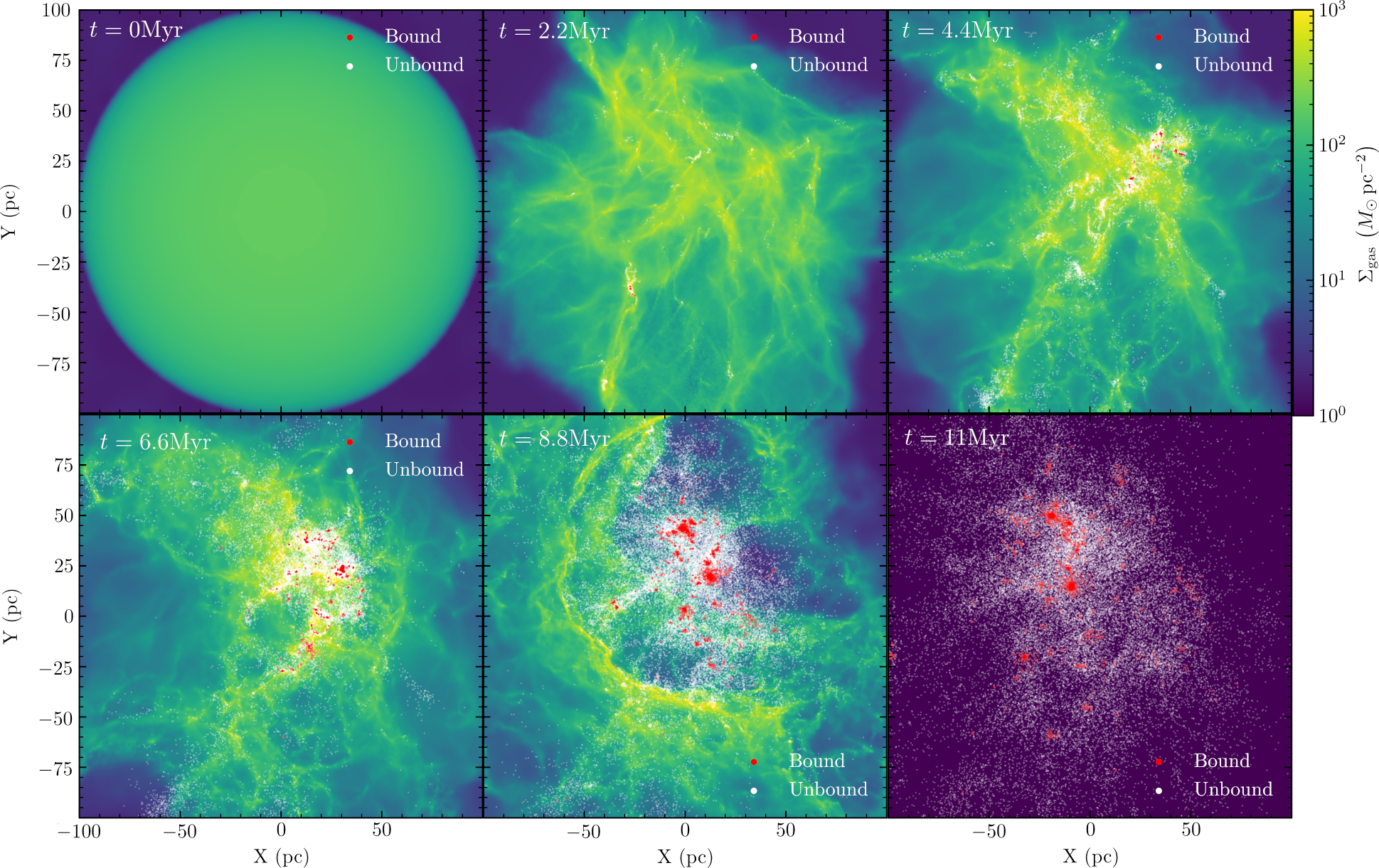}
\caption{State of the fiducial cloud model with $M=4\times 10^6 M_\odot$, $R=100\mathrm{pc}$, and random seed 1 at various points in its evolution (elapsed time given at the top left of each respective panel). Positions of star particles are plotted atop the gas surface density distribution. Star particles not assigned to any bound cluster are shown in white, and bound star particles in red. The GMC produces a highly sub-structured complex of both unbound and bound stellar mass, with most stellar mass in the unbound association in this case ($f_{\rm bound} \sim 10\%$). This is superfically similar to the configuration of observed complexes of newly-formed stars (see e.g. \citet{kuhn:2014}, Fig. 2).}
\label{fig:association}
\end{figure*}

All simulations follow the sequence of events that is typical in GMC simulations with a complete accounting of stellar feedback. The initial turbulent motions dissipate on a crossing time-scale \citep{stone:1998.dissipation}, inevitably leading to localized runaway collapse and eventually star formation. The SFR accelerates at first \citep{murray.chang:2015,lee:2015.gravoturbulence, lee:2016.gmc.eff}, but eventually levels off and begins to drop as feedback begins to evacuate the gas \citep{feldmann:2011,grudic:2018.gmc.sfe,li:2019.cfe}. Eventually, all gas is evacuated by feedback and star formation ceases entirely. This process, from start to finish, is illustrated for our fiducial cloud model, with mass $M_{\rm GMC}=4\times 10^6 M_\odot$ and $R_{\rm GMC}=\unit[100]{pc}$ (Figure \ref{fig:association}).
All clouds were evolved for two initial cloud free-fall times,
\begin{equation}
t_{\rm ff,0} = \frac{\pi}{2}\sqrt{\frac{R_{\rm GMC}^3}{G M_{\rm GMC}}},
\end{equation}
typically on the order of $10 \rm Myr$ for cloud parameters studied here. We generally find that star formation has almost ceased entirely by $\sim 1t_{\rm ff,0}$, and the central region is essentially gas-free by $2t_{\rm ff,0}$.

\subsection{Star formation efficiency}
\begin{figure}
\centering
\includegraphics[width=\columnwidth]{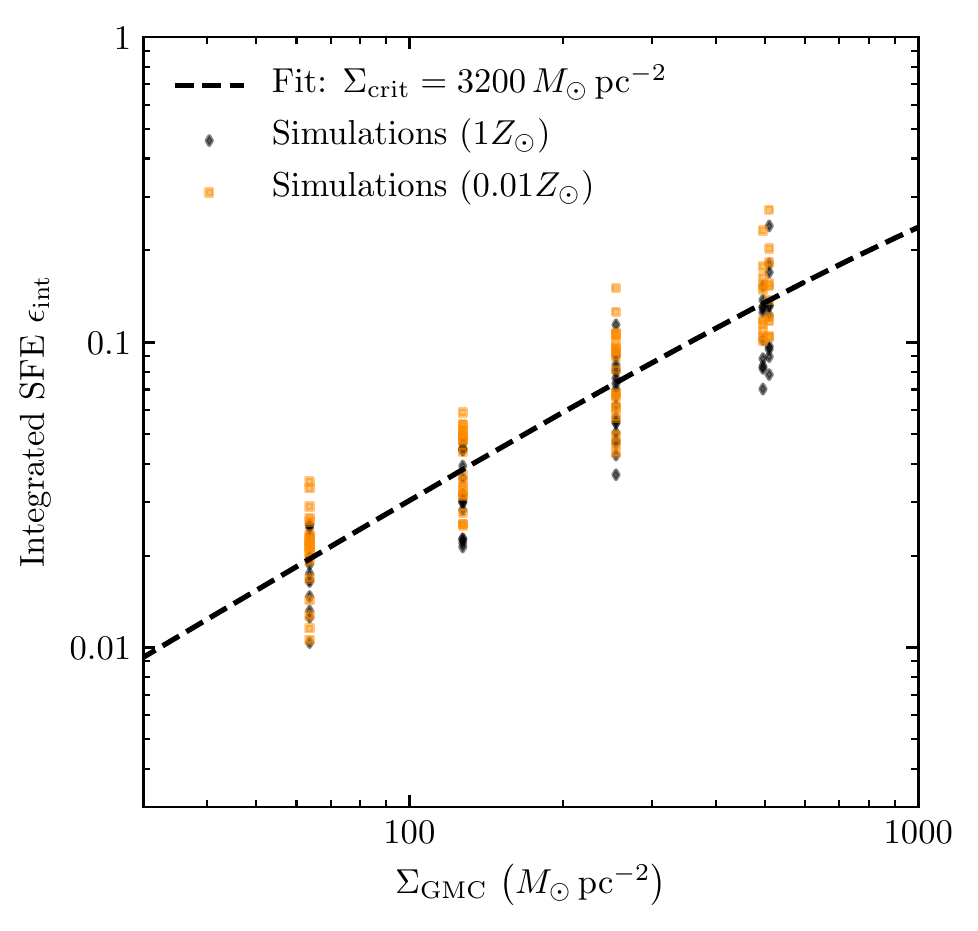}
\caption{Integrated star formation efficiency $\epsilon_{\rm int}=\frac{M_\star}{M_{\rm GMC}}$ as a function of the mean initial cloud surface density, $\Sigma_{\rm GMC}=\frac{M_{\rm GMC}}{\uppi R_{\rm GMC}^2}$. $\epsilon_{\rm int}$ scales roughly $\propto \Sigma_{\rm GMC}$, and is well-fit by Equation \ref{eq:sfefit} with $\Sigma_{\rm crit}=\unit[3200]{\mspc}$ and $\epsilon_{\rm int}^{\rm max}=0.68$, with $\unit[0.13]{dex}$ of residual scatter. This scatter is driven by cloud-to-cloud variations in the details of the initial turbulent gas flow, but is relatively small compared to the scatter in the bound fraction of star formation (e.g. Figure \ref{fig:CFE}).}
\label{fig:SFE}
\end{figure}
For this new simulation suite, which has different parameter space and initial cloud kinematics from the \citet{grudic:2018.gmc.sfe} suite, we repeat the excercise of determining the strongest predictor of the integrated SFE, $\epsilon_{\rm int}$, in terms of the initial cloud parameters. We again find that the cloud surface density $\Sigma_{\rm GMC}$ is the tightest predictor of $\epsilon_{\rm int}$, in agreement with the picture where $\epsilon_{\rm int}$ is primarily determined by the force balance of feedback and gravity within the cloud, which yields a dimensional scaling \citep{fall:2010.sf.eff.vs.surfacedensity,murray:2010.sfe.mw.gmc}.

We plot the relation between $\Sigma_{\rm GMC}$ and $\epsilon_{\rm int}$ in Figure \ref{fig:SFE}, and see that as before it scales roughly $\propto \Sigma_{\rm GMC}$, with a modest scatter of $\unit[0.1]{dex}$ between the different metallicities and statistical realizations of the cloud models, which we will neglect in all subsequent modeling. We perform an unweighted least-squares fit on $\log \epsilon_{\rm int}$ as a function of $\Sigma_{\rm GMC}$ as in \citet{grudic:2018.gmc.sfe}:

\begin{equation}
    \epsilon_{\rm int} = \left(\frac{1}{\epsilon_{\rm int}^{\rm max}} + \left(\frac{\Sigma_{\rm GMC}}{\Sigma_{\rm crit}}\right)^{-1}\right)^{-1},
    \label{eq:sfefit}
\end{equation}
and find best-fit parameters $\Sigma_{\rm crit}=\unit[3200]{\mspc}$ and $\epsilon_{\rm int}^{\rm max}=0.7$\footnote{We are hardly able to constrain the maximum SFE $\epsilon_{\rm int}^{\rm max}$ here due to our choice of parameter space, where the SFE does not exceed $25\%$, unlike \citep{grudic:2016.sfe} which surveyed much higher $\Sigma_{\rm GMC}$ and consequently higher $\epsilon_\mathrm{int}$.}. We therefore confirm that the SFE scaling formula of \citet{fall:2010.sf.eff.vs.surfacedensity} and \citet{grudic:2016.sfe} still holds for clouds whose internal motions are dominated by turbulence. We also find that our model with $M_{\rm GMC}=\unit[2\times 10^6]{M_\odot}$ and $R_{\rm GMC}=\unit[100]{pc}$ has a SFE of $1-3\%$, similar to \citet{grudic:2018.gmc.sfe} where we used rather different, ``pre-stirred" turbulent initial conditions from a driven turbulent box simulation. The details of how turbulence is initialized in isolated cloud simulations does not appear to materially affect the outcome of star formation when feedback is fully accounted for. Of course, both the pre-stirring and Gaussian random field methods are equally artificial, as the initial turbulent motions are not self-consistent with effects of gravity and feedback.

The relative robustness of the SFE to the different statistical realizations suggests that the star formation efficiency depends simply on the balance of feedback and gravity on the \textit{cloud} scale. Since the cloud bulk properties are essentially a controlled variable here, we accordingly see little variation from cloud to cloud of a given set of parameters.

\subsection{The nature of cluster formation}

The state of our fiducial cloud model after the end of star formation is plotted in Figure \ref{fig:association} panels 5 and 6, and is qualitatively representative of the result of all simulations in the suite. The result of star formation is a sub-structured, clustered configuration, similar to observed complexes of young stars \citep[e.g.][]{kuhn:2014,gouliermis:2018.review,ward.kruijssen:2019}. Some stars are in bound clusters, but distinct clustered substructures also exist that are not bound. The unbound component eventually disperses in roughly a cloud crossing time, while the respective bound components survive and virialize through violent relaxation.

All simulated clouds produce {\it some} mass in gravitationally-bound star clusters, although there are a few cases in the lowest-$\Sigma_{\rm GMC}$ runs where the only cluster masses fall below our resolution cut of 32 particles (which will serve as our resolution cut in this work, as this is roughly the stencil size over which the gravitational softening is adapted). Thus, in the simulations, the formation of bound versus unbound stellar systems are not different modes of star formation, but rather different aspects of the same underlying continuum, consistent with the hierarchical picture of star cluster formation \citep{kruijssen:2012.cluster.formation.efficiency}.

This alone does not rule out the monolithic picture: within this picture, the unbound stellar component at the end of star formation could simply consist of formally-bound clusters that have been disrupted by gas removal. However, we have verified that {\it final} bound fraction is of the same order (within a factor of $\sim 2$) as the fraction of stars that have {\it ever} belonged to {\it any} structure bound by stellar self-gravity. In other words, the unbound component is dominated by stars that have never had the chance to orbit within a cluster. Therefore, our simulation results are well-described by the hierarchical scenario.

\subsection{The bound fraction of star formation}
\label{sec:sims:results:cfe}
\begin{figure}
\includegraphics[width=\columnwidth]{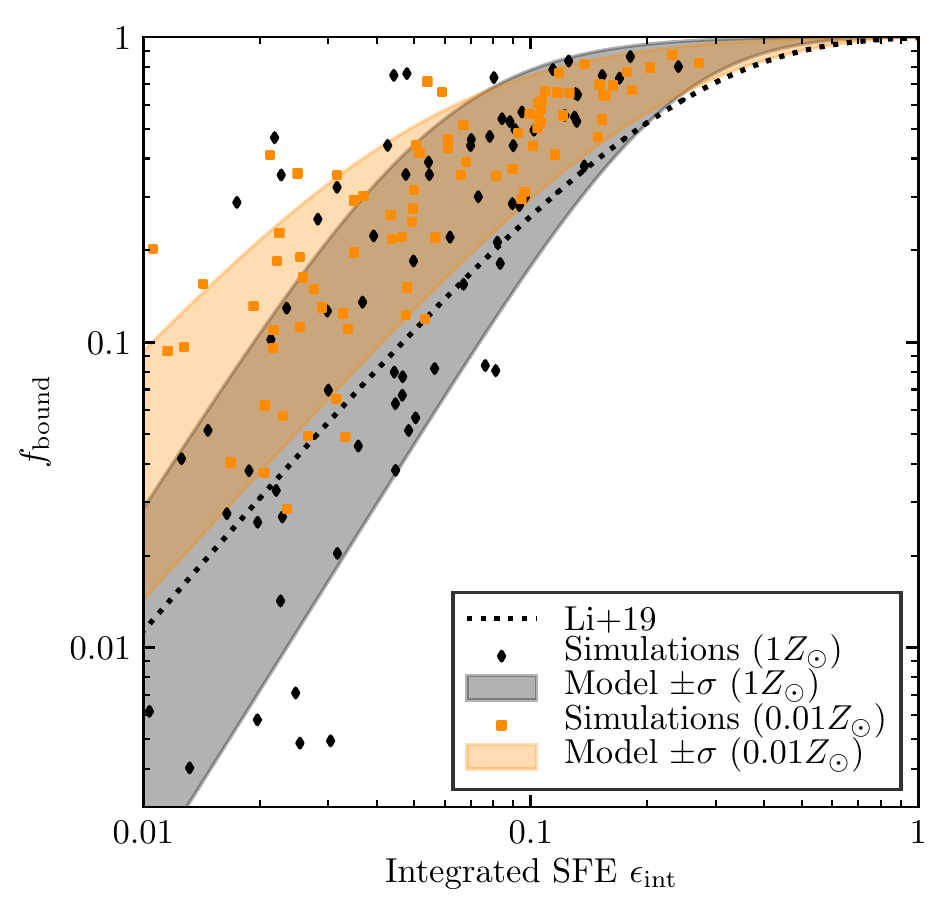}
\caption{Fraction of stars formed in bound clusters $f_{\rm bound}$ as a function of the integrated star formation efficiency $\epsilon_{\rm int}$. Points are individual simulations, shaded regions are the $\pm \sigma$ contours from the derived statistical model (\S\ref{section:model}), and the dotted line indicates the fit given by \citet{li:2019.cfe} to their simulation results. $f_{\rm bound}$ and $\epsilon_{\rm int}$ are correlated, but not generally equal, and with significant scatter from one turbulent realization to another at lower $\epsilon_{\rm int}$. $f_{\rm bound}$ saturates to $\sim 1$ at a SFE of $\sim 20\%$. It is also systematically greater at low metallicity due to the lack of strong OB winds.}
\label{fig:SFEvCFE}
\end{figure}

\begin{figure}
\centering
\includegraphics[width=\columnwidth]{{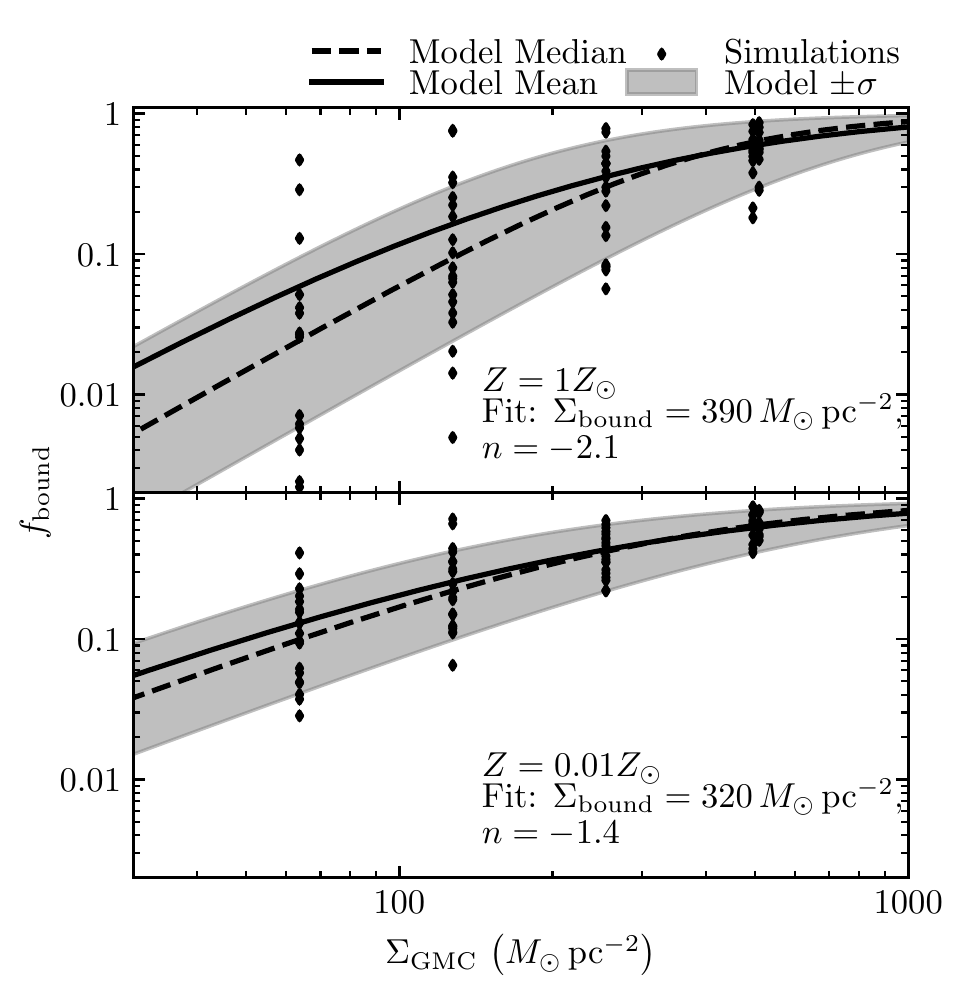}}
\caption{$f_{\rm bound}$, the fraction of stars in bound clusters at the end of star formation in the simulated GMC, as function of GMC surface density $\Sigma_{\rm GMC}$ for solar metallicity (\textit{top}) and $1\%$ solar metallicity (\textit{bottom}). We plot the result for each individual simulation (black dots). Unlike the SFE, the relation is notably metallicity-dependent. We perform different fits of equation \ref{eq:cfefit} for each respective metallicity, with best-fit parameters listed above - the solar-metallicity relation is a steeper function of $\Sigma_{\rm GMC}$, and reaches the $50\%$ mark at slightly greater surface density. The shaded regions indicate the $\pm \sigma$ region our model for the scatter in $f_{\rm bound}$ (Equations \ref{eq:sigmaprime} and \ref{eq:cfefit2})}.
\label{fig:CFE}
\end{figure}

At the end of each simulation, at time $2 t_{\rm ff,0}$, we ran the cluster-finding algorithm described in \S\ref{section:clusterfinding} to group the star particles in to bound star clusters. Across the entire suite, we identified 6181 bound clusters resolved by more than 32 star particles. Every cloud produced at least one bound cluster, and typically multiple ones. As expected from analytic work \citep{hills:1980,mathieu:1983}, N-body experiments \citep{tutukov:1978, lada:1984, kroupa:2001.orion.pleiades, baumgardt.kroupa:2007}, and recent star formation simulations with an idealized feedback model \citep{li:2019.cfe}, the fraction of stars remaining in bound clusters at the end of star formation is an increasing function of $\epsilon_{\rm int}$ (Fig. \ref{fig:SFEvCFE}). Also in agreement with previous numerical works, the bound fraction saturates to $\sim 1$ at considerably lower SFE than the $50\%$ required in the limit of fast gas evacuation in the classic \citet{hills:1980} derivation -- the typical $f_{\rm bound}$ is $50\%$ when $\epsilon_{\rm int} \sim 10\%$. Systematically higher $f_{\rm bound}$ is achieved at $1\%$ solar metallicity, which we will show in \S \ref{sec:physicstests} can be isolated to the lack of strong OB winds, whose mass loss rates scale $\propto Z^{0.7}$ \citep{vink:2001.ob.mass.loss}.

In Figure \ref{fig:SFEvCFE} we also plot the fit to the semi-analytic model derived in \citep{li:2019.cfe} that summarizes their simulation results. Our results are in reasonably good agreement with this model, modulo a factor of $\sim 2$ in the SFE where $f_\mathrm{bound} \rightarrow 1$. This is interesting because their $\Sigma_\mathrm{GMC}-\epsilon_\mathrm{int}$ relation was much higher than our Figure \ref{fig:SFE} for most of their simulations, and yet our simulations lie on nearly the same curve in the $\epsilon_\mathrm{int}-f_\mathrm{bound}$ plane. This suggests that this relation is robust to the details of hydro solvers (ie. {\small GIZMO} MHD vs. {\small AREPO} HD), and feedback model (realistic multi-channel vs. idealized local momentum injection).

The most striking result of Figure \ref{fig:SFEvCFE} is the large scatter in $f_\mathrm{bound}$ at fixed $\epsilon_{\rm int}$: the relation between the two quantities is not one-to-one, and must be modeled statistically. This is also readily seen in Figure \ref{fig:CFE}, where we plot $f_{\rm bound}$ as a function of $\Sigma_{\rm GMC}$. At fixed $\Sigma_{\rm GMC}$ (which is effectively fixed $\epsilon_{\rm int}$ as well), $f_{\rm bound}$ varies by as much as $\unit[2]{dex}$, particularly at lower $\Sigma_{\rm GMC}$. Surprisingly, there is at least one cloud at any given surface density that is able to form $>50\%$ of its stars in bound clusters. This result was anticipated by idealized N-body experiments of substructured cluster assembly \citep{smith:2011.cluster.assembly, smith:2013.cluster.assembly}.

Despite the large scatter, a clear scaling in the ${\it typical}$ bound fraction can be discerned. For each metallicity, we performed unweighted least-squares fits of $\log f_{\rm bound}$ to a generic saturating power-law in $\Sigma_{\rm GMC}$:
\begin{equation}
    f_{\rm bound} = \left( 1 + \left(\frac{\Sigma_{\rm bound}\left(Z\right)}{\Sigma_{\rm GMC}}\right)^{n\left(Z\right)}\right)^{-1},
    \label{eq:cfefit}
\end{equation}
where $\Sigma_{\rm bound}\left(Z\right)$ denotes the metallicity-dependent surface density at which $f_{\rm bound}=50\%$ and $n\left(Z\right)$ is metallicity-dependent power-law slope in the limit $\Sigma_{\rm GMC}<<\Sigma_{\rm bound}$. We found $\Sigma_{\rm bound}\left(Z_\odot\right)=\unit[390]{\mspc}$, $\Sigma_{\rm bound}\left(0.01Z_\odot\right)=\unit[330]{\mspc}$, $n\left(Z_\odot\right)=-2.1$, and $n\left(0.01Z_\odot\right)=-1.4$. 

We can model the stochastic variation in $f_{\rm bound}$ shown in Figure \ref{fig:CFE} by introducing a logarithmic variance in the ``effective" value of $\Sigma_{\rm GMC}$ that is plugged into Equation \ref{eq:cfefit}. The effective surface density $\Sigma'$ is distributed according to a log-normal distribution centered on the actual $\Sigma_{\rm GMC}$:
\begin{equation}
    P\left(\ln \Sigma' | \sigma_\mathrm{b}\right) = \frac{1}{\sqrt{2\pi \sigma_\mathrm{b}^2}}\exp\left(\frac{-\left(\ln \Sigma'-\ln \Sigma_{\rm GMC}\right)^2}{2 \sigma_\mathrm{b}^{2}}\right).
    \label{eq:sigmaprime}
\end{equation}
Once sampled from this distribution, the effective surface density is then plugged into Equation \ref{eq:cfefit}:
\begin{equation}
    f_{\rm bound} = \left( 1 + \left(\frac{\Sigma_{\rm bound}\left(Z\right)}{\Sigma'}\right)^{n\left(Z\right)}\right)^{-1}.
    \label{eq:cfefit2}
\end{equation}
The results of this model are plotted as the shaded regions in Figure \ref{fig:CFE}. We have found that the scatter $\Sigma_{\rm GMC}$-dependent scatter in $f_{\rm bound}$ is well-reproduced by the parameter $\sigma_b=0.7$, ie. the ``effective" surface density that matters for star cluster formation varies intrinsically by $\unit[0.3]{dex}$ from one cloud to another, due to the varying small-scale details of the cluster-forming gas flows. 

This procedure can also be repeated to account for the variations in $\epsilon_{\rm int}$ (Figure \ref{fig:SFE}, but we find that this is not as nearly important for obtaining a faithful description of the similation results as it is for $f_{\rm bound}$. First, $\epsilon_{\rm int}$ is not as steep a function of $\Sigma_{\rm GMC}$, so variations in the ``effective" surface density do not compound as drastically. Second, the maximum-likelihood value of $\sigma$ for modeling variations in $\epsilon_{\rm int}$ is less than $\nicefrac{1}{3}$ the value required for $f_{\rm bound}$. We therefore neglect the intrinsic scatter in $\epsilon_{int}$ in the results of this work.

\subsection{Mass distribution of bound star clusters}
\begin{figure}
    \includegraphics[width=\columnwidth]{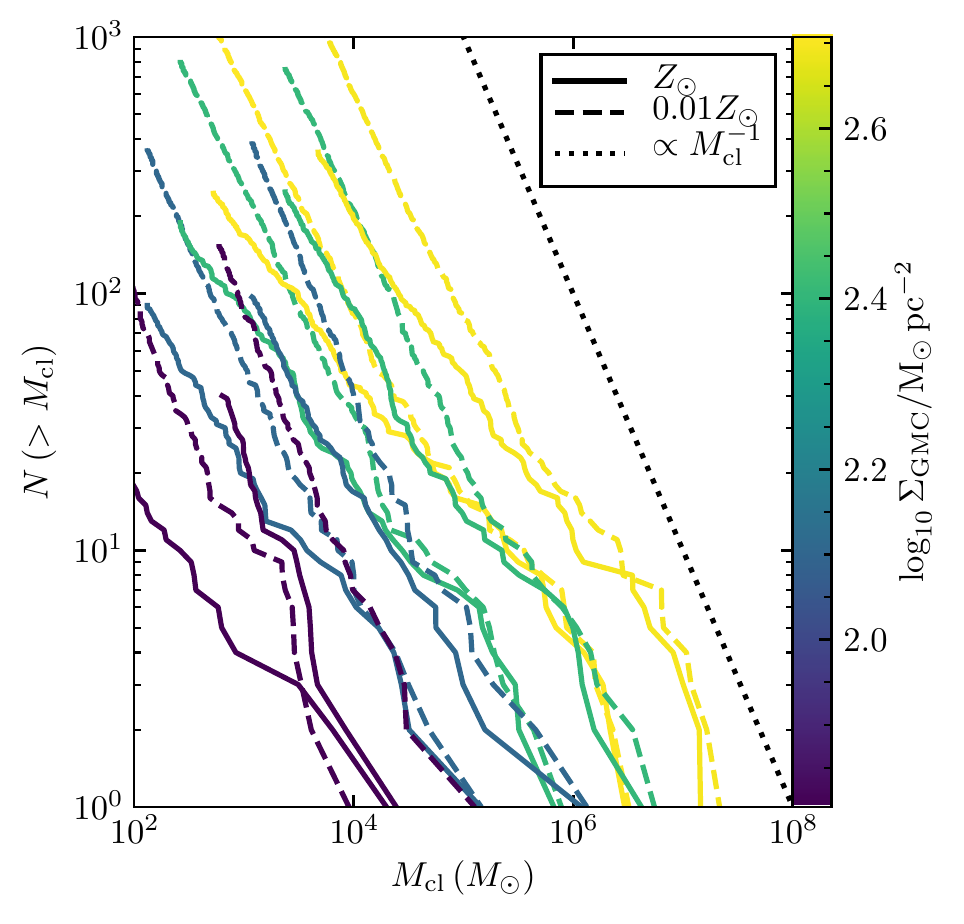}
    \caption{Cumulative mass functions of bound clusters produced by each cloud in parameter space, stacked over the ten random initial turbulent seedings. Curves are colour-coded according to the initial cloud surface density $\Sigma_{\rm GMC}$, and we differentiate between solar-metallicity (solid) and $1\%$ solar metallicity (dashed) cloud models. The dotted line shows the cumulative mass function expected from a typical star cluster mass function of the form $\dndm\propto M_{\rm cl}^{-2}$. See Figure \ref{fig:m83massfunc} for model predictions of the mass function in a real galaxy.}
    \label{fig:massfunc}
\end{figure}

\begin{figure}
    \centering
    \includegraphics[width=\columnwidth]{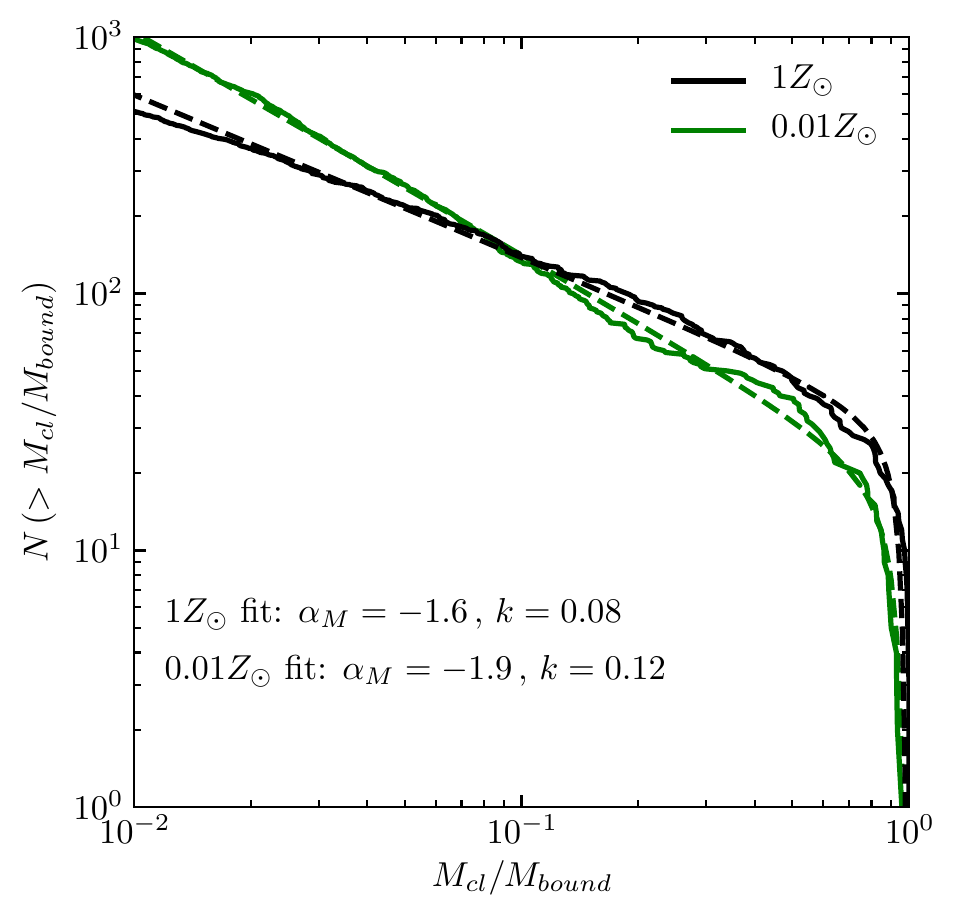}
    \caption{Cumulative distribution of bound star clusters in all simulations, relative to the total bound mass in the respective run. The total mass tends to be dominated by the massive primary cluster. We separate the distributions by metallicity, as the low-metallicity runs tend to have steeper mass functions. Fits to Equation \ref{eq:massfuncfit} are shown as dashed lines.}
    \label{fig:normmassfunc}
\end{figure}

As shown in Figure \ref{fig:association}, the simulated GMCs typically produce multiple bound star clusters of varying masses and sizes -- the mapping from clouds to clusters is not one-to-one. Rather, it must be understood in terms of a \textit{distribution} of cluster masses that emerges at the level of individual clouds. In general, we find that the mass distribution can be described in terms of a primary cluster that dominates (has $\sim \nicefrac{1}{10}-\nicefrac{1}{2}$) of the total mass in bound clusters, and a population of less-massive clusters whose masses are distributed according to a power-law. The mass distribution of bound clusters is robust to numerical resolution (Appendix \ref{section:appendix}).

In Figure \ref{fig:massfunc} we show the stacked cumulative mass functions from all realizations of a give set of cloud parameters -- ie., the mass distribution that would be observed if a galaxy formed clusters from many GMCs, but with uniform bulk properties. The distributions are all fairly top-heavy, due to the few particularly massive clusters in the population that constitute the primary cluster in each respective cloud. However, the asymptotic slopes of the mass functions toward lower masses tend to have power-law behaviour, often scaling roughly $\propto M_{\rm cl}^{-1}$, consistent with the mass distribution $\sim -2$ that is typically measured in young star cluster populations \citep[e.g.][]{adamo:2020.cluster.review}. The mass distributions from the solar metallicity clouds have preferentially shallower slopes, more consistent with a mass function of slope $\sim -1.6$. It should be noted that the mass distributions in Figure \ref{fig:massfunc} do not in themselves constitute predictions of a galactic star cluster mass distribution, as GMC bulk properties are not uniform in real galaxies. We perform a more-realistic synthesis of observable mass functions in \S \ref{section:model}.

The cloud-level mass functions can be better summarized when collapsed down in terms of the mass relative to the total bound mass, $M_{\rm cl}/M_{\rm bound}$. In Figure \ref{fig:normmassfunc} we plot the stacked relative mass functions for the two different metallicities, over all simulations. The cumulative distributions are strongly concave down (such that the PDF $\frac{\rm d N_{\rm cl}}{\rm d\log  M_{\rm cl}}$ is {\it peaked}) in the vicinity of $M_{\rm cl}/M_{\rm bound}$ because the primary tends to be so dominant, but they then level off to a power-law tail. We fit these cumulative distributions to the model
\begin{equation}
    N\left(>M_{\rm cl}\right) = N_0 \left(\frac{M_{\rm cl}}{M_{\rm bound}}\right)^{1-\alpha_{\rm M}\left(Z\right)} \exp\left(\frac{-k\left(Z\right)}{1-\frac{M_{\rm cl}}{M_{\rm bound}}}\right),
    \label{eq:massfuncfit}
\end{equation}
where $N_0$ is a normalization factor, $\alpha_{\rm M}\left(Z\right)$ is the slope of the asymptotic power-law mass function for small masses, and $k\left(Z\right)$ is a dimensionless shape parameter that models the heavy top end of the distribution. Again we form metallicity-dependent fits, finding $\alpha_{\rm M}\left(Z_\odot\right)=-1.6$, $k\left(Z_\odot\right)=0.08$, $\alpha_{\rm M}\left(0.01Z_\odot\right)=-1.9$, and $k\left(0.01Z_\odot\right)=0.12$. We therefore see that even controlling for GMC properties, a power-law star cluster mass distribution of between $-1.6$ and $-1.9$ emerges, due to the cluster multiplicity inherent in hierarchical star formation.

\subsection{Size-mass relation}\label{sec:sizemass}
\begin{figure}
    \centering
    \includegraphics[width=\columnwidth]{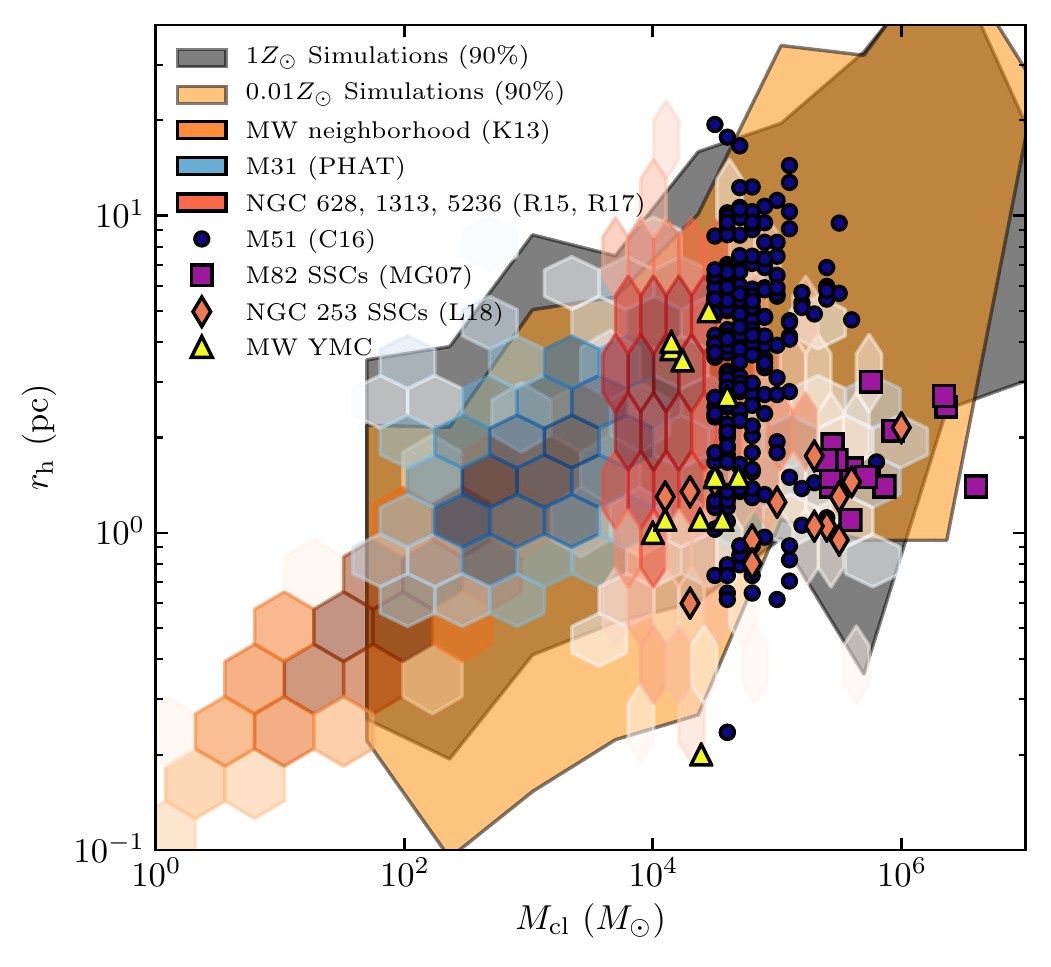}
    \caption{Size-mass relation of the star cluster catalogue extracted from the simulation suite, compared with observed catalogues. We plot the mass-binned contours containing $90\%$ of simulated clusters, with $Z_\odot$ runs and $0.01Z_\odot$ runs shown in grey and green respectively. Observational data include nearby ($<\unit[2]{kpc}$) clusters in the Milky Way (\citealt{karchenko:2013.mwg.clusters}, K13), Milky Way young massive clusters from the compilation of \citet{krumholz:2018.star.cluster.review} (MW YMC), and young clusters in NGCs 628, 1313, and 5236 (\citealt{ryon:2015.m83.clusters, ryon:2017.ymc.profiles}, R15, R17), M31 (\citealt{johnson:2012.phat, fouesneau:2014.phat}, PHAT), M82 (\citealt{mccrady:m82.sscs}, MG07), and NGC 254 (\citealt{leroy:2018.ssc}, L18). This figure is largely reproduced from \citet{krumholz:2018.star.cluster.review} with permission, code and data courtesy of Mark Krumholz.}
    \label{fig:massradius}
\end{figure}

We measured the clusters' projected half-light radii $r_{\rm h}$ integrated along the $z$ axis in the simulation coordinates, centred on the point of maximum intensity. We caution that unlike cluster masses, our cluster sizes do exhibit some systematic dependence upon numerical resolution (see Appendix \ref{section:appendix}) in the range of mass resolutions we simulated: the clusters were systematically smaller and denser at higher resolution. This may be an unavoidable artifact of our collisionless approximation, as at some point stellar dynamical effects should be important for limiting the phase-space density of stars. Therefore we do not rule out that our results here are subject to numerical effects. Nevertheless, for modeling purposes, we will find that the model derived from our simulations succeeds at reproducing star clsuter sizes{\it a posteriori} in Section \ref{section:M83}, so we still present the basic simulation results here.

The values of $r_\mathrm{h}$ we obtained are summarized in Figure \ref{fig:massradius}, where we compare the range of sizes of simulated star clusters with various populations of clusters in the local Universe. We find that the simulated clusters generally do lie in the space of mass and size of observed young star clusters, with less-massive clusters tending to be smaller and vice versa, but with such a weak dependence that no trend at all would likely be seen in a cluster catalogue spanning $<\unit[2]{dex}$ in mass. As with real star cluster populations, the most salient feature of the simulated mass-size relation is the considerable {\it dispersion} at fixed mass. Across the entire simulated catalogue, we find a dispersion in cluster size of $\unit[0.5]{dex}$, which remains roughly constant with mass. 

We have tried fitting $r_{\rm h}$ to a general power-law of the form $\propto M_{\rm GMC}^{\alpha_1} \Sigma_{\rm GMC}^{\alpha_2} Z^{\alpha_3} M_{\rm cl}^{\alpha_4}$, and the logarithmic least-squares best fit is best predicted by

\begin{equation}
    r_{\rm h} = \unit[3]{pc}\left(\frac{M_{\rm GMC}}{\unit[10^6]{M_\odot}}\right)^\frac{1}{5}\left( \frac{\Sigma_{\rm GMC}}{\unit[100]{\mspc}} \right)^{-1} \left(\frac{Z}{Z_\odot}\right)^\frac{1}{10} \left(\frac{M_{\rm cl}}{\unit[10^4]{M_\odot}}\right)^\frac{1}{3},
    \label{eq:sizefit}
\end{equation}
with $\pm \unit[0.4]{dex}$ of residual scatter that is not driven by variations in the quantities considered above, which is well-approximated by a log-normal distribution. Thus, although the intrinsic scatter is considerable, we do find a mass-size relation that is set by the cloud properties. Neglecting the very weak dependencies upon cloud mass and metallicity (assuming $10^6M_\odot$ and $Z_\odot$), the size-mass relation lies along lines of constant 3D stellar density, with the 3D density of clusters set by $\Sigma_{\rm GMC}$:

\begin{equation}
    \rho_\mathrm{eff} \equiv \frac{3 M_{\rm cl}}{8\pi r_{\rm h}^3} \approx \unit[44]{M_\odot\,\mathrm{pc}^{-3}} \left(\frac{\Sigma_{\rm GMC}}{\unit[100]{\mspc}}\right)^3 \pm \unit[1.1]{dex}
    \label{eq:sizefit2}
\end{equation}

Assuming that $\Sigma_\mathrm{GMC}$ is an increasing function of the mean galactic $\Sigma_\mathrm{gas}$, such an underlying relation might explain the observed characteristic 3D density of young star clusters in local spiral galaxies (the line $\log \rho = 2$ in Figure \ref{fig:massradius}), where most stars form in clouds with $\sim \unit[50-100]{\mspc}$.  It may also explain the relative compactness of star clusters formed in starburst galaxies like M82  for their mass \citep{mccrady:m82.sscs} compared to young clusters in typical spiral galaxies: the central region of M82 has a mean gas surface density of $\Sigma_\mathrm{gas}\sim \unit[500]{\mspc}$ \citep{weiss:2001.m82}, ie. $\sim 5$ times greater than the typical GMC surface density in galaxies where clusters lie along the typical size-mass relation, and hence the typical star cluster density is $5^3\sim \unit[2]{dex}$ greater. However this remains fairly speculative, as a proper numerical comparison would require $\Sigma_\mathrm{GMC}$ to be known, and likely also some accounting for cluster size evolution \citep[e.g.][]{choksi:2019.cluster.size}.


\subsubsection{The roles of different feedback mechanisms}
\label{sec:physicstests}
\begin{table}
    \centering
    \begin{tabular}{c|c|c|c}
    Run     &  $\epsilon_{\rm int}$ & $f_{\rm bound}$ & $M_{\rm cl,max}$ ($M_\odot$)\\
    \hline
    Standard, $Z_\odot$ & 4.6\% & 6.7\% &$5 \times 10^3$\\
    \hline
    Standard, $0.01Z_\odot$ & 4.3\% & 22\% & $10^4$\\
    \hline
    No winds & 4.6\% & 18\% & $1.3 \times 10^4$\\
    \hline
    No radiation & 10.4\% & 70.4\% & $2.7 \times 10^5$  \\
    \hline
    No SNe &4.6\% & 5.3\% & $6\times 10^3$\\
    \hline
    No feedback & 30\% + & 90\%+ & $3.2 \times 10^5+$ \\
    \end{tabular}
    \caption{$\epsilon_{\rm int}$, $f_{\rm bound}$, and the maximum bound cluster mass in test runs that turn off various subsets of the feedback physics included in the standard suite, run for the fiducial cloud with $M_{\rm GMC}=\unit[4\times10^6]{M_\odot}$ and $R_{\rm GMC}=\unit[100]{pc}$ (for a single turbulence realization). Results of the `No feedback' model are given with a `+' because this model was only run for half as long as the others, and at this time these quantities were still rising.}
    \label{table:physicstests}
\end{table}
To determine which specific feedback mechanisms are responsible for the setting the various quantities presented in this section, we ran a series of simulations on our fiducial cloud model ($M_{\rm GMC}=\unit[4\times10^6]{M_\odot}$, $R_{\rm GMC}=\unit[100]{pc}$, and $Z=Z_\odot$) in which we varied the feedback physics included for a single turbulence realization. Specifically, we tried switching off stellar winds, radiation, supernovae, and all feedback mechanisms in turn. We summarize the results of this experiment in Table \ref{table:physicstests}. 

Neglecting feedback altogether results in runaway collapse and very high $\epsilon_{\rm int}$, $f_{\rm bound}$, and star cluster mass. We did not run the no-feedback run past $t_{\rm ff,0}$ due to the computational expense of integrating the extremely dense star clusters that formed, and at this time $30\%$ of the cloud mass had been converted into stars, with no sign of stopping.

Neglecting radiation increased $\epsilon_{\rm int}$ from $5\%$ to $10\%$, and $f_{\rm bound}$ from $7\%$ to $70\%$. Radiative feedback is therefore apparently crucial in moderating star cluster formation, and also plays the dominant role in setting the cloud-scale SFE, although stellar winds and SNe are still able to moderate star formation somewhat.

The results of the run neglecting SNe are nearly identical to the standard run, so we find that SNe are practically irrelevant to both the cloud-scale SFE and the formation of bound clusters in this region of parameter space. They are unable to moderate star formation on the scale of cluster-forming clumps because the clusters generally form over much shorter time-scales than the $\sim\unit[3]{Myr}$ that it takes for the first SNe to go off.

Neglecting stellar winds did not change $\epsilon_{\rm int}$ at all. However, the bound fraction and maximum cluster mass of the model with no stellar winds at {\it solar} metallicity were very close to those of the standard $0.01Z_\odot$ run. We are therefore able to isolate the metallicity dependence of $f_\mathrm{bound}$ shown in \S\ref{sec:sims:results:cfe} to the effective absence of stellar winds at low metallicity, as this is the only large metallicity dependence of feedback that we model. It is remarkable that stellar winds should affect cluster formation so drastically while leaving the cloud-scale SFE unaltered. In the dense ($\Sigma_{\rm gas} > \unit[10^3]{\mspc}$) clumps where individual clusters form, it is expected that both radiative feedback and the momentum-injecting component of stellar wind feedback are inefficient \citep{fall:2010.sf.eff.vs.surfacedensity}. By contrast, a hot stellar wind bubble that has not had a chance to vent may behave more in the regime of energy-conserving feedback, which is more efficient than momentum-conserving feedback on small scales: $\epsilon_{\rm int} \propto \Sigma^\frac{3}{2} R^\frac{1}{2}$, versus being $\propto \Sigma$ for momentum-conserving feedback \citep{fall:2010.sf.eff.vs.surfacedensity}.

To summarize, having some type of stellar feedback is crucial for setting both the cloud-scale SFE and the bound fraction of star formation. Radiative feedback is the most important, but stellar winds can be a uniquely efficient feedback mechanism on small scales, potentially affecting the outcome of individual star cluster formation while having modest effects upon the cloud-scale SFE.

\section{Statistical Model: Mapping Clouds to Clusters} \label{section:model}
\subsection{Harmonizing cloud parameters}
Equipped with general results for $\epsilon_{\rm int}$, $f_{\rm bound}$, and the star cluster size and mass distributions from the previous section, we are nearly ready to construct a statistical model that is able to reproduce our simulation results for any set of GMC parameters. But first, if the model is to be used on observational data, or clouds in a simulated galaxy, we must address some ambiguities in the model inputs: the cloud bulk parameters. For the purposes of the present work we will set aside the observational uncertainties about the interpretation of CO emission as a mass tracer, and assume that the CO to $H_\mathrm{2}$ conversion factor $X_\mathrm{CO}$ is known.

For real clouds, there is some ambiguity about what the proper cloud size $R_{\rm GMC}$, and more crucially surface density $\Sigma_{\rm GMC}$ to use is, as real clouds are not uniform spheres. For a general mass distribution $\rho\left(\mathbf{x}\right)$, we will define $R_{\rm GMC}$ as the radius of a sphere of equal moment of inertia:
\begin{equation}
    R_{\rm GMC} \equiv  \sqrt{\frac{5}{3M_{\rm GMC}} \int \rho\left(x\right) r^2\,\mathrm{d}^{3}\mathbf{x}},
    \label{eq:RGMC}
\end{equation}
where $r$ is the distance from the cloud centre of mass. This trivially reduces to our definition for a spherical top-hat distribution.

A common definition of the effective radius used in GMC catalogues is the root mean square of the intensity-weighted second moments of the 2D CO intensity \citep[e.g.][]{freeman:2017.m83.gmcs}. For a Gaussian cloud model, assuming CO intensity maps directly onto surface density, this definition is a factor of $\sqrt{3}$ less than the definition in Equation \ref{eq:RGMC}, and is a factor of $\sqrt{\frac{15}{2\pi}}$ less for a uniform sphere model. The two conversion factors are nearly equal -- we will adopt the latter in \S\ref{section:M83}.

Another common definition of the effective radius of a cloud is the radius of a circle with area equal to the pixels that the cloud occupies. This is more problematic for us, because it ultimately depends on the specific intensity cut that is used to define the cloud boundary. For this reason, we advise caution if applying this model to data that uses this definition -- it is not obvious that the effective radius provided is actually characteristic of the mass distribution of the cloud. A decently representative value of $\Sigma_{\rm GMC}$ is crucial for the model, because this affects both $\epsilon_{\rm int}$ and $f_\mathrm{bound}$, so cluster masses are doubly sensitive to it. Thus modest errors in $R_{\rm GMC}$ can compound into major errors in star cluster properties.

\begin{table}
    \centering
    \begin{tabular}{c|c|c|c|c}
    Parameter &  Used in & Affects & $Z_\odot$ value & $0.01Z_\odot$ value \\
    \hline
    $\Sigma_{\rm crit} $ & Eq. \ref{eq:sfefit} & SFE & $\unit[3200]{\mspc}$ &  $\unit[3200]{\mspc}$ \\
    \hline
    $\epsilon_{\rm int}^{\rm max}$ & Eq. \ref{eq:sfefit} & SFE & 0.8 & 0.8 \\
    \hline
    $\Sigma_{\rm bound}$ & Eq. \ref{eq:cfefit2} & $f_{\rm bound}$ & $ \unit[390]{\mspc}$ & $\unit[330]{\mspc} $\\
    \hline
    n & Eq. \ref{eq:cfefit2} & $f_{\rm bound}$ & 2 & 1.4 \\
    \hline
    $\sigma_{b}$ & Eq \ref{eq:sigmaprime} & $f_{\rm bound}$ & 0.7 & 0.7 \\    
    \hline
    $\alpha_{M}$ & Eq \ref{eq:massfuncfit} & Mass function & -1.6 & -1.9 \\
    \hline
    $k$ & Eq \ref{eq:massfuncfit} & Mass function & 0.08 & 0.13 \\
    \end{tabular}
    \caption{Summary of model parameters for mapping GMCs onto star cluster populations. None of these are free parameters: the are calibrated to reproduce the star cluster statistics of the simulation results in Section \ref{section:sims:results}.}
    \label{table:params}
\end{table}

\subsection{Algorithm}
Having resolved the ambiguity in cloud size, we are now equipped with the cloud bulk parameters that are the inputs to the model: its mass $M_{\rm GMC}$, radius $R_{\rm GMC}$, mean surface density $\Sigma_{\rm GMC}=\frac{M_{\rm GMC}}{\pi R_{\rm GMC}^2}$, and metallicity $Z$. The mapping from clouds to clusters then proceeds as follows:
\begin{enumerate}
    \item Compute the metallicity-dependent model parameters provided by the simulations: $\Sigma_{\rm bound}\left(Z\right)$, $n\left(Z\right)$ (Equation \ref{eq:cfefit2}), $\alpha_{\rm M}\left(Z\right)$, and $k\left(Z\right)$ (Equation \ref{eq:massfuncfit}). For an arbitrary metallicity, we use a linear interpolant in $\log Z$, using the values provided at $0.01Z_\odot$ and $Z_\odot$ in Table \ref{table:params}. For any parameter $p\left(Z\right)$:
    \begin{equation}
        p\left(Z\right) = \frac{\log_{10}\left(Z/Z_\odot\right)+2}{2} p\left(Z_\odot\right) - \frac{\log_{10}\left(Z/Z_\odot\right)}{2} p\left(0.01Z_\odot\right)
    \end{equation}
    \item Compute the SFE (Equation \ref{eq:sfefit}) and the total stellar mass formed:
    \begin{equation}
        M_\star = \epsilon_{\rm int} M_{\rm GMC} 
    \end{equation}
    \item Of this total stellar mass that forms, compute the fraction $f_{\rm bound}$ of this stellar mass in bound clusters with Equations \ref{eq:sigmaprime} and \ref{eq:cfefit2}. The total mass in bound clusters is then:
    \begin{equation}
        M_{\rm bound} = M_{\rm GMC} \epsilon_{\rm int} f_{\rm bound}.
    \end{equation}
    The mass that is not in bound clusters constitutes the unbound association component of the stellar population formed in the cloud.
    \item Sample the relative cluster masses $M_{\rm cl}/M_{\rm bound}$ from the cloud-level mass distribution (Equation \ref{eq:massfuncfit}), until the sum of the masses exceeds $M_{\rm bound}$. Reject the final cluster if it commits a lesser mass conservation error than keeping it.
    \item Sample the cluster half-mass radii according to a log-normal size distribution with median given by Eq. \ref{eq:sizefit} and with variance $0.4\rm dex$.
\end{enumerate}
Finally, using the results of \citet{grudic:2017}, we can also model the specific shapes of the star cluster density profiles. Young star clusters are generally well-fit by the \citet{Elson:1987.ymc.profile} density profile:
\begin{equation}
    \rho \left( r \right) = \rho_0 \left(1 + \frac{r^2}{a^2}\right)^{-\frac{\gamma+1}{2}},
\end{equation}
where $a$ is a scale radius, related to the effective radius by
\begin{equation}
    a = \frac{r_{\rm h}}{\sqrt{2^\frac{2}{\gamma-2}-1}}
\end{equation}
and $\gamma$ is the power-law slope of the outer surface density profile of the cluster. We found in \citet{grudic:2017} that $\gamma$ has a universal distribution in observed and simulation star cluster populations, that is seemingly uncorrelated with any other cluster property, but is apparently set during the star formation process.  For our synthetic cluster population, we sample $\gamma$ randomly using the following fit to the universal CDF on the interval $\gamma \in [2,10]$:
\begin{equation}
    N\left(< \gamma\right) = 1.064 \left(\frac{\gamma -2}{\gamma - 0.8}\right)^{0.54}.
\end{equation}

\subsection{Monte Carlo experiments}
\label{section:montecarlo}
\begin{figure*}
    \centering
    \includegraphics[width=\textwidth]{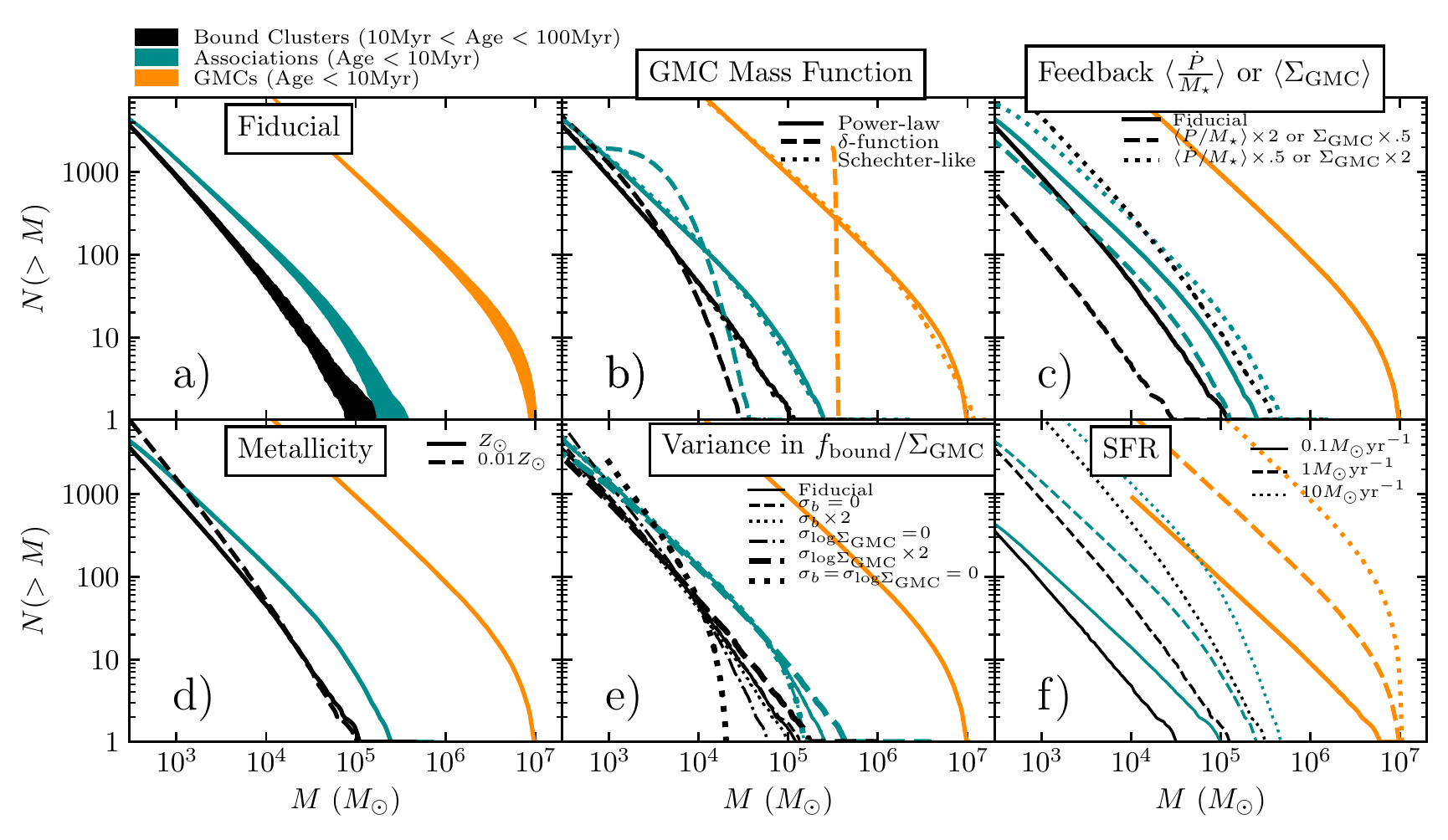}
    \caption{Results of Monte Carlo sampling experiments of our cluster formation model (\S \ref{section:model}) in a one-zone galaxy model with a continuously-forming GMC population of a certain size and mass distribution. We map the properties of GMCs (orange) onto the cumulative mass functions of young stellar associations (blue) and bound clusters (black), showing distributions in 10Myr age windows for GMCs and associations, and 90Myr windows for bound clusters, corresponding to typical age ranges considered in observations.  Contours show $\pm \sigma$ intervals over 100 time windows, curves show median values. {\bf a)} Fiducial model: a $\propto M^{-2}$ power-law GMC mass distribution truncated at $10^4$ and $10^{7}M_\odot$, with $\langle \Sigma_\mathrm{GMC}\rangle=50M_\odot\,\mathrm{pc}^{-2}$, $\sigma_{\log\Sigma_\mathrm{GMC}}=0.3\rm dex$, sampled at a rate so that the galactic SFR is $1M_\odot\,\rm yr^{-1}$. Contours show $\pm \sigma$ intervals over 100 realizations. {\bf b)} Varying the GMC mass distribution, for the fiducial power-law, Schecter-like, and Dirac $\delta$-function forms. {\bf c)} Varying the strength of stellar feedback $\langle \dot{P}/M_\star\rangle$, roughly equivalent to inversely varying $\langle \Sigma_\mathrm{GMC}\rangle$. {\bf d)} Varying metallicity. {\bf e)} Varying $\sigma_b$ and $\sigma_{\log \Sigma_\mathrm{GMC}}$, which affect the intrinsic variance in $f_\mathrm{bound}$ and $\epsilon_\mathrm{int}$. {\bf f)} Varying the galactic SFR.}
    \label{fig:massfunc_model}
\end{figure*}

To build some intuition for the predictions of the model for typical galactic conditions, we now consider its predictions for the star cluster population arising in a simple one-zone galactic ISM model described by the following properties, with respective fiducial values:
\begin{enumerate}
    \item Star formation rate: SFR (fiducial value $1M_\odot\,\mathrm{yr}^{-1}$).
    \item Distribution of GMC formation rates in GMC mass: $\frac{\mathrm{d}N_{\rm GMC}}{\mathrm{d} t\,\mathrm{d} M_{\rm GMC}}$. For our fiducial model, we take a power-law distribution $\frac{\mathrm{d}N_{\rm GMC}}{\mathrm{d} t\,\mathrm{d} M_{\rm GMC}} \propto M_{\rm GMC}^{-2}$ with lower limit $M_{\rm GMC,min}=10^3M_\odot$ and upper limit $M_{\rm GMC, max} = 10^7 M_\odot$. 
    \item Distribution of GMC surface densities, which we model as a log-normal distribution with mean $\langle \Sigma_{\rm GMC} \rangle =  \unit[50]{M_\odot\,pc^{-2}}$ and dispersion $\sigma_{\log \Sigma_{\rm GMC}} = \unit[0.3]{dex}$, similar to what is found in GMC catalogues in our and nearby galaxies \citep{miville:2017.gmcs, freeman:2017.m83.gmcs,faesi.lada:2018.gmcs,sun:2018.gmcs}, and in agreement with galactic-scale simulations with stellar feedback and resolved ISM structure \citep{hopkins:fb.ism.prop,guszejnov:2020.gmcs}.
\end{enumerate}

We synthesize star cluster populations from our model by sampling a sufficiently large sample of $N_{\rm GMC}$ GMC masses and surface densities from the formation-rate mass function and $\Sigma_{\rm GMC}$ distributions respectively, determining the total stellar mass $M_\star$ formed by plugging these into Equation \ref{eq:sfefit}, and implicitly determining the times between individual GMC star formation episodes as $\Delta t = M_\star / \left(\mathrm{SFR} \,N_{\rm GMC}\right)$. We assume a constant spacing in time $\Delta t$, but very similar results were obtained from a random Poisson process with mean GMC formation time-scale equal to this $\Delta t$.

In Figure \ref{fig:massfunc_model} we plot simple mock observations of the mass CDFs for GMCs, stellar associations, and bound star clusters in our model galaxy (taking stellar associations to be the {\it entire} stellar content formed by a GMC, bound or unbound). We show cumulative GMC and stellar association populations formed in $\unit[10]{Myr}$ windows (effectively assuming a $\unit[10]{Myr}$ lifetime, \citealt{kruijssen:2019.ngc300,chevance:2020.gmcs}), and clusters in the age range $\unit[10-100]{Myr}$ windows for bound clusters, similar to the age bins typically assumed to consist mainly of bound clusters in observations \citep[e.g.][]{adamo:2015.m83.clusters,johnson:2016.cluster.formation.efficiency,messa:2018.m51}. 

Panel a) shows the relationship between the mock-observed mass functions of GMCs, associations, and bound clusters for our fiducial model. The stellar association mass function is essentially the GMC mass function shifted downward by a factor of the SFE, and convolved with a log-normal due to the variance in $\Sigma_\mathrm{GMC}$ and hence $\epsilon_\mathrm{int}$. The mass function of bound clusters is somewhat steeper than that of GMCs or associations, and has less resemblance in shape, with a less-obvious upper truncation at our fiducial SFR. The maximum cluster mass is less than the maximum association mass, but only by a factor of $\sim 2$, considerably greater than would be assumed by multiplying the maximum stellar association mass by the mean galactic $f_\mathrm{bound}$ of $\sim 10\%$.

In Figure \ref{fig:massfunc_model} panel b) we vary the assumed GMC mass distribution, comparing our fiducial truncated power-law with a \citet{schechter:1976}-like $\dndm \propto M_\mathrm{cl}^{-2} \exp\left(-M_\mathrm{cl}/M^\ast\right)$ form and a Dirac $\delta$-function, all normalized to have equal mass-weighted median GMC mass. The truncated power-law and Schechter-like models are difficult to distinguish for all 3 mass functions, illustrating the importance of statistical rigor when attempting to distinguish between these models in observations \citep[e.g.][]{johnson:2017.m31.massfunction,mok:2019,adamo:2020.cluster.review}. The $\delta$-function gives the ``impulse response" of our model, from which any mass function can be constructed via synthesis. The resulting stellar association mass function is log-normal due to the log-normal $\Sigma_\mathrm{GMC}$ and $\epsilon_\mathrm{int}$ distributions, while the bound cluster mass function exhibits a low-mass power-law tail imprinted by the GMC-level mass function.

In Figure \ref{fig:massfunc_model} panel c) we vary the strength of feedback, as expressed by the specific momentum injection rate from a young stellar population $\langle \dot{P}/M_\star\rangle$. We achieve this by simply re-scaling the {\it surface densities} plugged into the model: according to the standard dimensional argument for the scaling of SFE with surface density \citet{fall:2010.sf.eff.vs.surfacedensity}, these are equivalent. Because the SFE-$f_\mathrm{bound}$ relation (Figure \ref{fig:SFEvCFE}) is robust to the specifics of feedback, this rescaling procedure should also model the consequences of varying $\langle \dot{P}/M_\star\rangle$ for $f_\mathrm{bound}$. As expected from Equation \ref{eq:sfefit}, varying $\langle \dot{P}/M_\star\rangle$ for $f_\mathrm{bound}$ by a factor of 2 simply rescales the masses of stellar associations inversely. However the masses of bound clusters are {\it doubly} sensitive, varying by a factor of $\sim 4$, because the variation in $f_\mathrm{bound}$ compounds with the variation in $\epsilon_{int}$. Cluster masses are therefore highly sensitive to the strength of stellar feedback.

The effect of varying metallicity in shown in Figure \ref{fig:massfunc_model} panel d) is subtle: although we found that $f_\mathrm{bound}$ is systematically greater at lower metallicity (Figures \ref{fig:CFE}-\ref{fig:SFEvCFE}), most of this extra mass shows up in the more bottom-heavy tail of the mass distribution, leaving the maximum cluster mass virtually unaffected.

Panel e) shows the sensitivity of the model to cloud-to-cloud variance in $\epsilon_\mathrm{int}$ and $f_\mathrm{bound}$, as driven by variance in $\Sigma_\mathrm{GMC}$ or the intrinsic variance parametrized by $\sigma_\mathrm{b}$. Varying $\sigma_\mathrm{b}$ alone at the factor of 2 level has very subtle effects, but varying $\sigma_{\log \Sigma_\mathrm{GMC}}$ clearly affects both the maximum stellar association and bound cluster masses. Setting both variances to 0 reduces the maximum cluster mass by nearly an order of magnitude: the maximum cluster mass can then be never be greater than $f_\mathrm{bound}\left(\Sigma_\mathrm{GMC}\right)\epsilon_\mathrm{int}\left(\Sigma_\mathrm{GMC}\right) M_\mathrm{GMC,max}$, with the respective efficiency factor given by Equations \ref{eq:sfefit} and \ref{eq:cfefit} amounting to $\sim 10^{-3}$, much less than the typical factor of $\sim 10^{-2}$ separating the largest cluster mass in a given galaxy from the largest GMC mass.

Finally, in Figure \ref{fig:massfunc_model} panel f) we vary the SFR assumed in the model. This is equivalent to sampling more GMCs and their resulting associations and bound clusters. The GMC mass distribution's upper cutoff becomes very apparent at higher SFR, while it becomes difficult to constraint for lower SFR. Because sampling more clouds samples makes rare, highly-efficient events more likely to occur in a given time window, the maximum masses of stellar associations and clusters in a given age window both scale with the SFR. The run with SFR is approximately $\propto \mathrm{SFR}^\frac{1}{2}$, similar to the observed relation between the brightest cluster mass and the galactic SFR \citep{bastian:2008.cfe}.

\section{Comparison with Observations} \label{section:M83}
\begin{figure}
    \centering
    \includegraphics[width=\columnwidth]{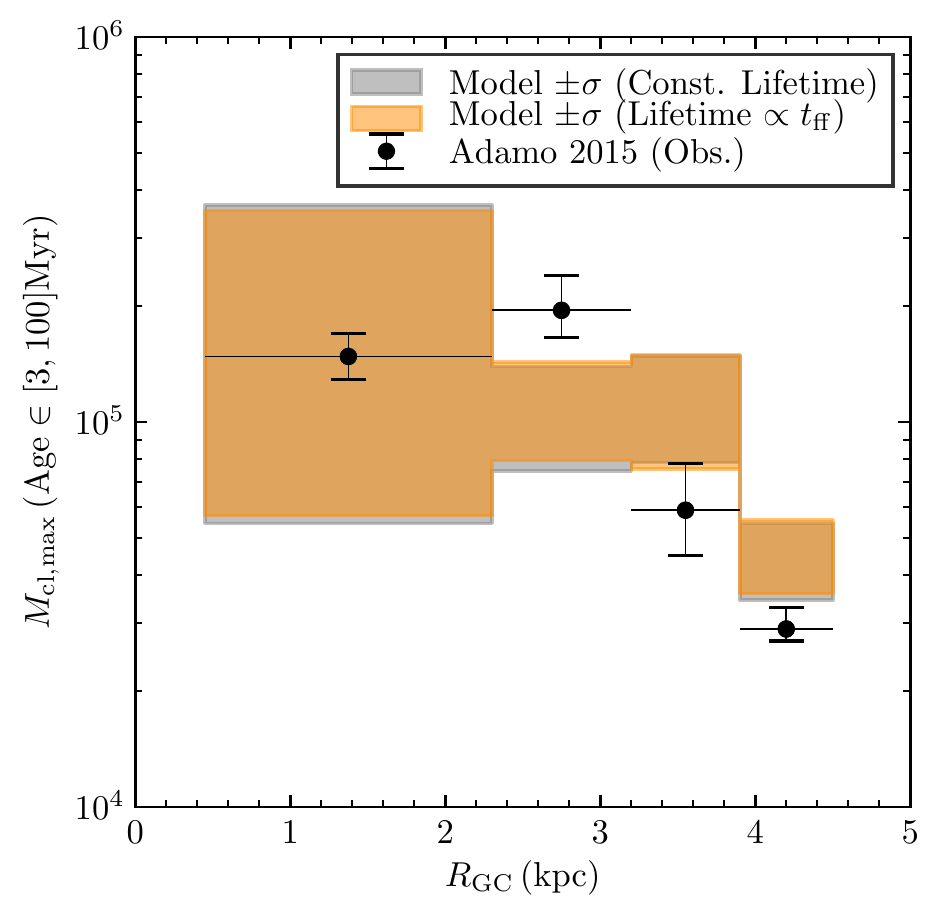}
    \caption{Comparison of the maximum bound star cluster mass from our Monte Carlo simulation of M83's cluster population with data from the \citet{adamo:2015.m83.clusters} star cluster catalogue, in four different radial bins and $97 \rm Myr$ age windows (see \S\ref{section:M83} for details). Shaded regions show the $\pm \sigma$ intervals over all observation times, assuming either constant GMC lifetimes (grey) or $\propto t_\mathrm{ff}$ lifetimes (orange) (\S\ref{section:M83:sampling}). } 
    \label{fig:m83mmax}
\end{figure}
\begin{figure}
\includegraphics[width=\columnwidth]{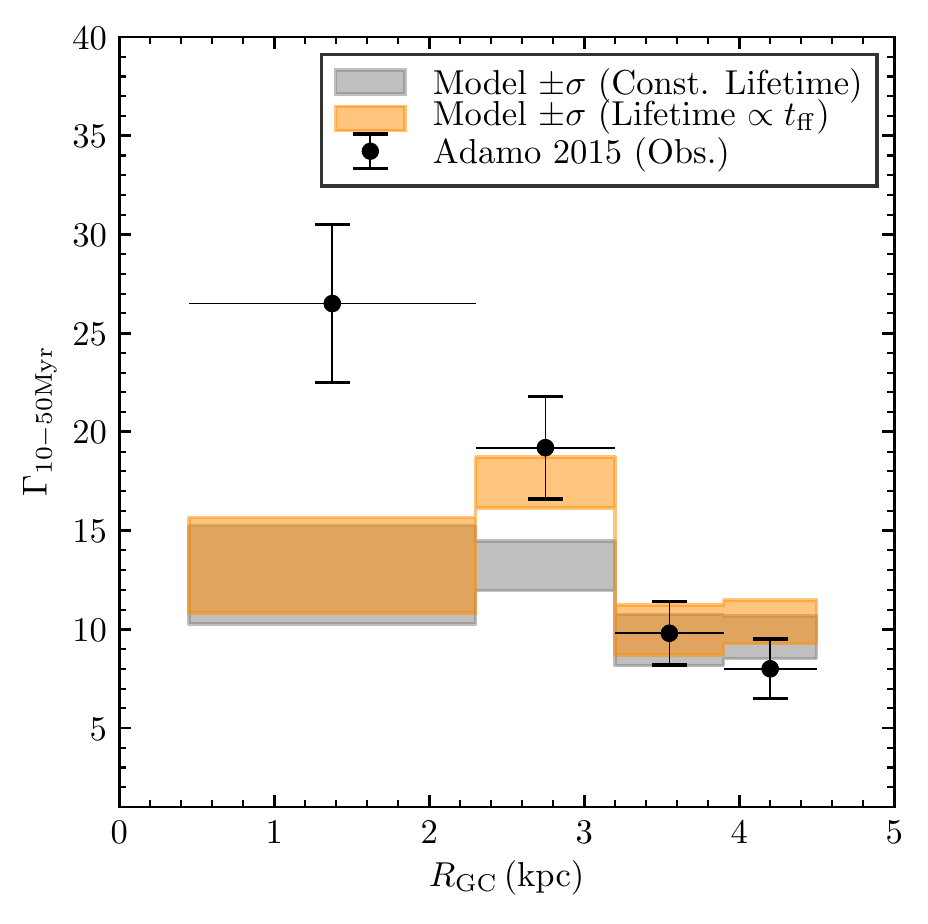}
\caption{Comparison the predicted bound cluster formation efficiency $\Gamma_\mathrm{10-50\mathrm{Myr}}$ in our Monte Carlo simulation of M83's cluster population with data from the \citet{adamo:2015.m83.clusters} star cluster catalogue, in four different radial bins and $40 \rm Myr$ age windows (see \S\ref{section:M83} for details). Shaded regions show the $\pm \sigma$ intervals over observation times, assuming either constant GMC lifetimes (grey) or $\propto t_\mathrm{ff}$ lifetimes (orange) (\S\ref{section:M83:sampling}).}
\label{fig:m83gamma}
\end{figure}
\begin{figure}
    \centering
    \includegraphics[width=\columnwidth]{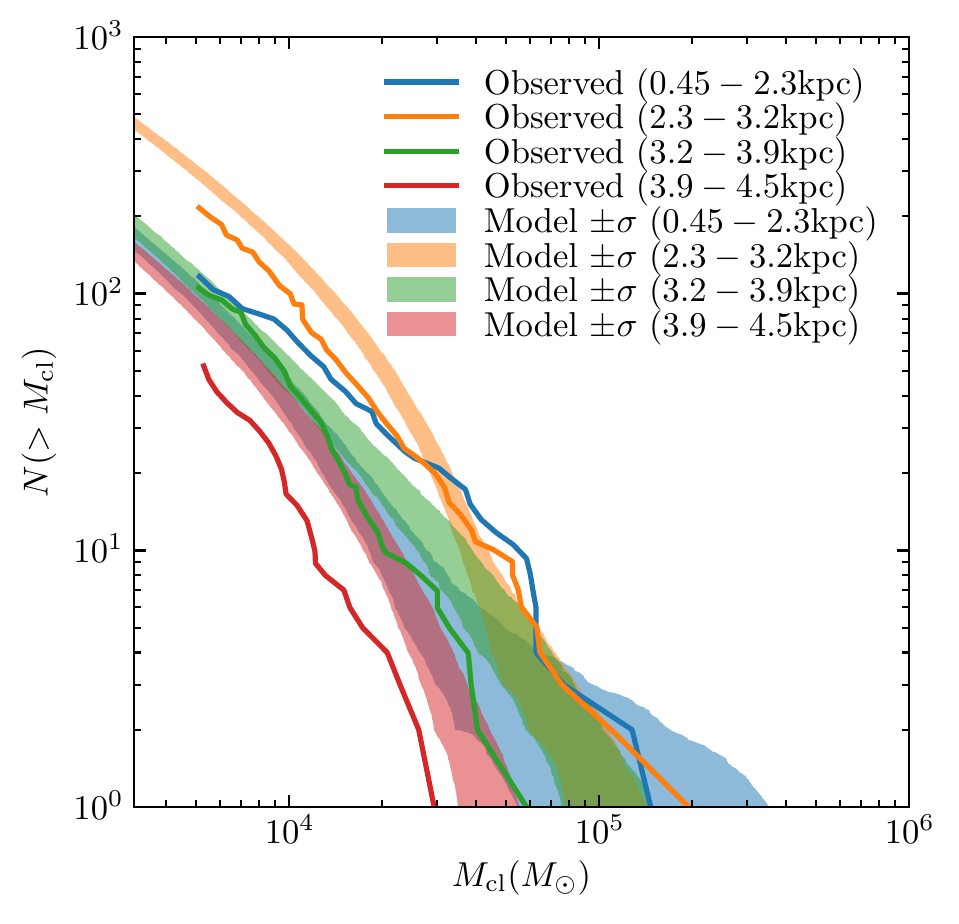}
    \caption{Predicted cumulative distributions of cluster masses of our Monte Carlo simulation of the M83 cluster population, compared with \citet{adamo:2015.m83.clusters} in 4 radial bins and $97 \rm Myr$ windows, assuming GMC lifetimes scale $\propto t_\mathrm{ff}$. Shaded regions show the $\pm \sigma$ contours over all observation times of the model.}
    \label{fig:m83massfunc}
\end{figure}
\begin{figure}
    \centering
    \includegraphics[width=\columnwidth]{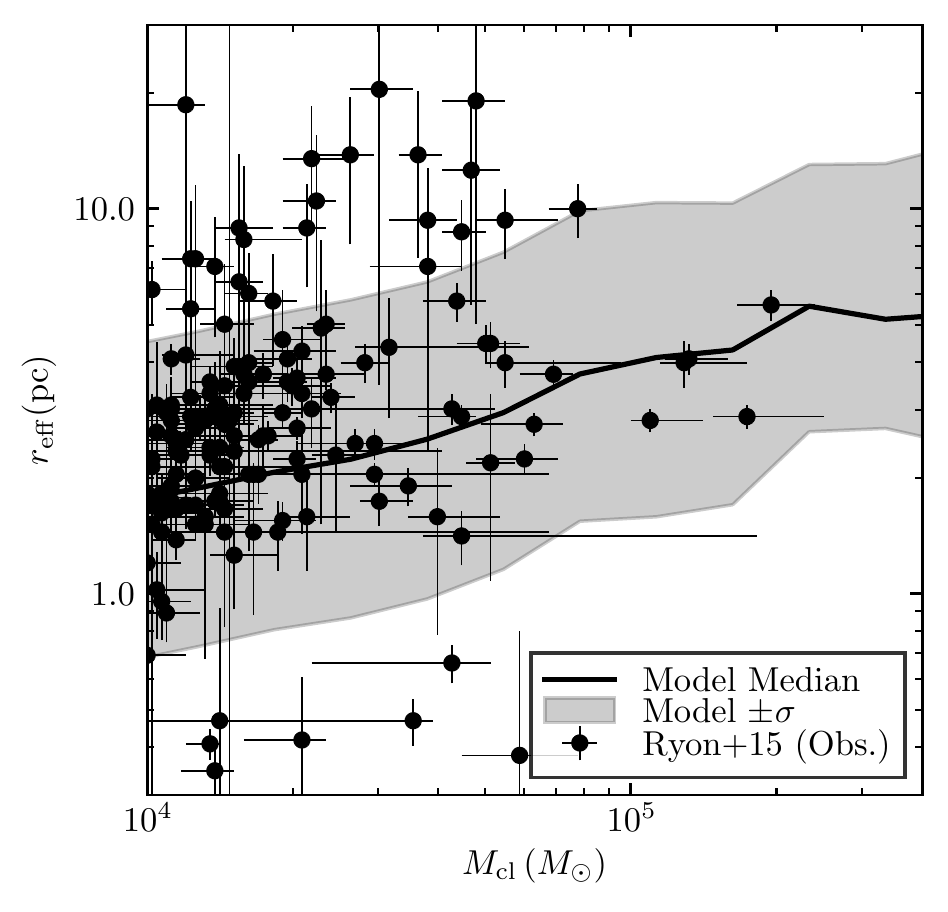}
    \caption{Predicted galactic size-mass relation of clusters in our Monte Carlo simulation of the M83 star cluster population, compared with data from \citep{ryon:2015.m83.clusters} for clusters in the age range $10-300\mathrm{Myr}$. The curve indicates the median value over all observation times, and the shaded region indicates the $\pm \sigma$ quantiles. Bars on data points indicate $\pm \sigma$ uncertainties.}
    \label{fig:m83sizefunc}
\end{figure}
We will now test the model described in the previous section by synthesizing mock star cluster catalogues in M83 from the properties of its GMCs. We use the \citet{freeman:2017.m83.gmcs} catalogue, taking $M_{\rm GMC}$ to be the mass inferred from the clouds' CO luminosity, and correcting the effective radii by a factor of $\sqrt{\frac{15}{2\pi}}$ to obtain $R_{\rm GMC}$, as defined in Equation \ref{eq:RGMC}. We assume a radially-dependent metallicty, with a central metallicity of $+0.2\mathrm{dex}$ with a metallicity gradient of $-0.04\mathrm{dex}\,\mathrm{pc}^{-1}$ \citep{hernandez:2019.m83.metallicity}.  We bin the GMCs and clusters by galactocentric radius as in \citet{adamo:2015.m83.clusters}, with bin edges $0.45$, $2.3$, $3.2$, $3.9$, and $4.5$ $\mathrm{kpc}$. The galactic SFR is $1.2 M_\odot$, and we use the respective binned SFRs given in \citet{adamo:2015.m83.clusters}. 

\subsection{Sampling procedure}
\label{section:M83:sampling}
To synthesize cluster populations, we simply repeat the Monte Carlo procedure in Section \ref{section:montecarlo} using the observed properties of the GMCs instead of sampling from an assumed distribution in the mass-size plane. We are equipped with the observed distribution of {\it currently present} and {\it observable} GMCs, so to obtain $\frac{\mathrm{d}N_{\rm GMC}}{\mathrm{d} t\,\mathrm{d} M_{\rm GMC}}$ we must apply some re-sampling procedure to account for cloud lifetimes and for incompleteness. For incompleteness, we weight the sampling probabilities of each cloud inversely to the mass-dependent completeness fraction given in \citet{freeman:2017.m83.gmcs}. To account for cloud lifetimes, we consider two possible simple assumptions. Applying no additional re-weighting models a constant cloud lifetime, which gives $\frac{\mathrm{d}N_{\rm GMC}}{\mathrm{d} t\,\mathrm{d} M_{\rm GMC}}$ the same shape as the observed present mass distribution. Scaling the respective sampling weights $\propto t_\mathrm{ff}^{-1}=\frac{\uppi}{2} \sqrt{\frac{R_\mathrm{GMC}^3}{G M_\mathrm{GMC}}}$ models the case where the cloud lifetime is $\propto t_\mathrm{ff}$. Present observations can hardly differentiate the two pictures (within a given galaxy), and the latter assumption has physical motivation, so we adopt the latter as our preferred model. However we still present basic results for both cases in Figures \ref{fig:m83gamma} and \ref{fig:m83mmax}, and find that this ambiguity hardly affects our results in practice.

\subsection{Caveats and approximations}
Our procedure for inferring $\frac{\mathrm{d}N_{\rm GMC}}{\mathrm{d} t\,\mathrm{d} M_{\rm GMC}}$ from the data and comparing the projected cluster population with observations has two major caveats. We predict cluster formation in presently-observed GMCs but compare these clusters with a catalogue of clusters formed as much as $100\mathrm{Myr}$ ago, so we expect agreement only if the SFR and ISM conditions in the respective radial of M83 have remained steady over at least that time-scale, which is not guaranteed. The comparison also requires that the dynamical evolution and disruption of the clusters has been negligible over this time-scale, ie. the observed masses and sizes have not deviated significantly from their masses and sizes at formation, which are the quantities predicted by the model. Stellar mass loss alone implies that this is not strictly true for any cluster: a bound cluster in isolation will decrease in mass and increase in radius over time. The effects of internal dynamical relaxation and the galactic tidal field upon the cluster will further compound mass loss and and radius increase. However, we neglect these processes here under the assumption that their effects on observables are small over the $\sim 100\mathrm{Myr}$ time-scales we consider.

\subsection{Results}

First, we predict the maximum cluster mass expected to form in each region of the galaxy, observing the maximum cluster mass present in each $\unit[97]{ Myr}$ window in the Monte Carlo simulation of each radial bin. In Figure \ref{fig:m83mmax} we compare the $\pm \sigma$ intervals of the Monte Carlo results with the \citet{adamo:2015.m83.clusters} catalogue, for the two different assumptions about cloud lifetimes outlined in \ref{section:M83:sampling}. We find that the most massive cluster observed in a $\unit[97]{Myr}$ age window can vary considerably over time, by $\sim 0.5\mathrm{dex}$ in the 3 outermost bins and by nearly $1\mathrm{dex}$ in the innermost bin. In all cases our results are within $2\sigma$ of observed masses, with no systematic residual trend, so our predicted maximum cluster mass is in reasonably good agreement with observations. The assumption about cloud lifetimes hardly affects this statistic at all, because the few most extreme clouds that produce the most extreme clusters are generally getting sampled regardless of the sampling scheme.

Next, in Figure \ref{fig:m83gamma} we plot the mass fraction of star formation in bound clusters in $40 \rm{Myr}$ windows, $\Gamma_{10-50\mathrm{Myr}}$, to compare with values measured in \citet{adamo:2015.m83.clusters}.  $\Gamma_{10-50\mathrm{Myr}}$ also has intrinsic variation from window to window of $0.2-0.3\mathrm{dex}$. This variation is much greater than the intrinsic variation of the SFR within the Monte Carlo model -- rather, it is driven by the varying respective $f_\mathrm{bound}$ of the GMCs sampled. In a realistic scenario where the SFR {\it is} varying intrinsically on kpc scales by an appreciable amount, this variation in $\Gamma_{10-50\mathrm{Myr}}$ would be even greater. Hence the amount of variation we predict is really a lower bound. With the exception of the innermost bin, the model predictions are $1\sigma$-compatible with observations with no systematic residual trend, so the model is mostly in very good agreement with observations. 

The largest discrepancy is in the innermost ($0.5-2.3\rm kpc$) bin, in which we underestimate $\Gamma$ by a factor of $\sim 2$: to match observations, our model would require greater values $\Sigma_\mathrm{GMC}$ to increase $f_\mathrm{bound}$ at the cloud level, and while there is a weak gradient in $\Sigma_\mathrm{GMC}$ in M83, it is very weak outside of the galactic center $<0.5\rm kpc$. Note that our innermost bin has a galactic dynamical time at most $t_\mathrm{dyn}\left(2.3\rm kpc\right) \sim \Omega\left(2.3\rm kpc\right)^{-1} \sim 15\rm Myr$ \citep{lundgren:2004.m83}, shorter than the $40 \rm Myr$ age range of the observed cluster sample. As such, in this instance it is quite possible that that presently-observed GMCs are not representative of those GMCs that formed the presently-observed clusters, as the region has experienced roughly an orbit's worth of evolution.

In Figure \ref{fig:m83massfunc} we plot the mass distributions in $\unit[97]{Myr}$ windows to compare with those plotted in \citet{adamo:2015.m83.clusters}, under the assumption that the GMC lifetime scales $\propto t_\mathrm{ff}$. In all radial bins the mass functions are generally within $1\sigma$ agreement with observations, modulo the uncertainties detailed in \S\ref{section:M83:sampling}. Since we have neglected mass loss and disruption, our results are consistent with a picture where these effects upon the cluster mass function are small over $\sim 100\mathrm{Myr}$ time-scales in M83.

Finally, we plot the modeled size-mass relation for clusters in the $10-300\rm Myr$ age bin in Figure \ref{fig:m83sizefunc}, comparing with the data provided by \citet{ryon:2015.m83.clusters}. Although our cloud-level size-mass relation gives $r_\mathrm{h} \propto M_\mathrm{cl}^{1/3}$ for a given set of cloud properties, the result of convolving the model over the GMC size-mass distribution is slightly flatter, with slope $\sim \nicefrac{1}{4}$, which is consistent with the slopes fitted in \citet{ryon:2015.m83.clusters}. We find an intrinsic scatter in the size-mass relation of $\sim 0.4 \mathrm{dex}$, which is mainly driven by the intrinsic scatter arising from the star formation process in our model (\S\ref{sec:sizemass}). Although we have neglected the effects of cluster evolution, we anticipate two main effects. First, the cluster sizes will become systematically larger with age due to mass loss and dynamical relaxation, so the normalization of the relation will increase. Second, clusters that initially formed too puffy will be subject to disruption in the galactic environment, while clusters that initially formed too compact will undergo dynamical evolution over shorter time-scales, causing them to puff up \citep{kruijssen:2011.natural.selection}. The net result will be a gradual reduction in the scatter of the size-mass relation.

\section{Discussion}
\label{sec:discussion}

Having developed and validated a model for the formation of stellar associations and bound star clusters from GMCs, we now discuss various implications of the model, the interpretation of various phenomena within this framework, and compare the framework with other works.

\subsection{The importance of stellar feedback in cluster formation}
According to our simulations, stellar feedback that can moderate star formation in GMCs appreciably is an essential ingredient in any model attempting to explain the masses of young, massive star clusters in the local Universe. Without feedback, our model would convert $>50\%$ of GMC mass into star cluster mass, resulting in star cluster masses far in excess of what is observed. 

Our model does assume that the GMCs in question are gravitationally bound, and in principle an alternative model to feedback-moderated star formation in GMCs is that GMCs are not gravitationally bound. However, the majority of observed massive ($>10^6M_\odot$) GMCs do have virial parameters that are consistent with gravitational boundedness \citep{bolatto:2008.gmc.properties, kauffmann:2013.cloud.virial.parameters, heyer:2015.gmcs, rice:2016.gmcs, miville:2017.gmcs}. In particular, the clouds in the \citet{freeman:2017.m83.gmcs} catalogue we have focused on in this work tend to have virial parameters consistent with boundedness, so a lack of boundedness cannot explain their low SFEs. 


\subsection{The importance of stochasticity in $f_\mathrm{bound}$}
We have shown that star formation efficiency is not a one-to-one predictor of $f_{\rm bound}$ (Figure \ref{fig:SFEvCFE}). Moreover we find that there is no cloud bulk property that predicts $f_{\rm bound}$ better than $\Sigma_{\rm gas}$, which does so rather loosely, motivating our statistical approach. This suggests that the bulk properties of a GMC do not contain sufficient information to predict the detailed properties of its star clusters in a deterministic sense. 

This is intuitive if one considers the large dynamic range separating the scale of clusters ($\sim \unit[1]{pc}$) and clouds. The SFE will depend upon the balance of feedback and gravity on the scale of the cloud, and thus has relatively little scatter (Figure \ref{fig:SFE}). Meanwhile, $f_{\rm bound}$ depends on the details of the small-scale cluster-forming gas flow, which can be decoupled from the cloud-scale properties.

\subsection{The initial mass function of star clusters} 
\label{sec:discussion:cmf}
\subsubsection{Slope}
The mass distribution of star clusters is the most fundamental statistic of a star cluster population. It has been measured in various local galaxies, with the typical finding that it is well-fit by a power-law of the form 
\begin{equation}
    \dndm \propto M_{\rm cl}^{\alpha_{\rm M}},
\end{equation}
with $\alpha_{\rm M}$ being typically $\sim -2$, with some, but relatively little variation when measured across entire galaxies \citep{chandar:2015.cfe,krumholz:2018.star.cluster.review,adamo:2020.cluster.review}. The simplest explanation for this mass function is that GMCs also have a mass function of this form. Although GMC mass functions are typically measured to be top-heavy (ie. slope shallower than $-2$), it can be argued that the GMC lifetime is likely mass-dependent ($t_{\rm ff}\propto M^\frac{1}{4}$ at fixed $\Sigma_{\rm GMC}$), and thus the distribution of {\it cloud formation rates}, which longer-lived clusters should trace, is steeper \citep{fall:2010.sf.eff.vs.surfacedensity}, and hence the cluster mass function is inherited from the GMC mass function, assuming a constant SFE and no cloud-to-cluster multiplicity.

Our Monte Carlo experiments with a power-law GMC mass function in \S\ref{section:montecarlo} show that this logic does hold approximately for the mapping from the GMC mass function to {\it stellar associations}, as the stellar association mass function is roughly the GMC mass function convolved with a log-normal peaked at the typical GMC star formation efficiency, which is mass-independent at fixed $\Sigma_\mathrm{GMC}$ within our model. 

However, we find in \S\ref{section:montecarlo} that the mass function of bound star clusters generally comes out somewhat steeper than $-2$ ($\sim -2.4)$. While some galaxies do have steeper mass function slopes (e.g. M83, \S\ref{section:M83}), $-2$ {\it does} appear to be the typical value obtained in galactic-scale mass function fits \citep{krumholz:2018.star.cluster.review}. Mass-dependent disruption or mass loss would tend to flatten the mass function if it disproportionately affects low-mass clusters, which is the general result for {\it all} proposed evolutionary processes except for dynamical friction (which affects more massive clusters). There is evidence for such such age-dependent flattening in M83 and M51 \citep{bastian:2012.m83.clusters,messa:2018.m51}. Meanwhile, in M31, where the age function implies negligible disruption over $\sim 100\mathrm{Myr}$ time-scales \citep{johnson:2017.m31.massfunction}, the mass function is steep ($\sim -2.5$).

In general, our model suggests that the slope of star cluster mass functions is only universal insofar as GMC mass functions and star cluster mass loss and disruption are universal, and can change as a function of cluster age.


\subsubsection{High-mass cut-off}
In certain instances where it has been possible to get good statistics on star clusters within a certain localized region of a galaxy \citep{adamo:2015.m83.clusters, johnson:2017.m31.massfunction}, some evidence has been found for a ``Schechter-like'' truncation in the mass function, ie.
\begin{equation}
    \dndm \propto M_\mathrm{cl}^{\alpha_{\rm M}} \exp\left(\frac{-M_{\rm cl}}{M^\star}\right),
\end{equation}

This may reflect some important characteristic physical scale encoded in the structure of the ISM, in star formation physics, or in galactic dynamics. According to our model, a bound fraction of $\sim 1$ is theoretically possible for any cloud, and so the absolute maximum bound cluster mass that can form is simply
\begin{equation}
    M_{\rm cl,max} = \epsilon_{\rm int,max} M_{\rm GMC,max},
\end{equation}
where the maximum integrated SFE $\epsilon_{\rm int,max}$ depends upon the maximum $\Sigma_\mathrm{GMC}$ as given by Equation \ref{eq:sfefit}. Thus, interpreted in this manner, a truncation in the cluster mass function is the result of a truncation in the GMC mass function \citep{kruijssen:2014.cluster.formation}. It has been proposed that this truncation is set by the Toomre mass, the maximum mass that can collapse against galactic shear (and possibly feedback) \citep[e.g.][]{hopkins:2012.excursion.set, reinacampos:2017.cluster.model}, and this picture agrees well with observations in M83 \citep{freeman:2017.m83.gmcs}.

\subsubsection{Low-mass cut-off or shallowing} \label{sec:minclustermass}

Some low-mass cutoff or shallowing in the mass function of bound clusters must exist, if observed galactic mass functions with slopes $\leq-2$ in the observed range are to contain finite overall mass. This is difficult to constrain directly in nearby galaxies, as it requires knowledge of two uncertain incompleteness corrections: observationally, catalogues are typically incomplete in the mass range below a few $ 10^3M_\odot$ \citep{adamo:2017.legus}, and physically, low-mass clusters are also those most subject to disruption and mass loss in the galactic environment. Despite these difficulties, the low-mass regime is an important ingredient for modeling star cluster populations \citep[e.g.][]{pfeffer:2018.emosaics}, and may well depend sensitively upon star formation and feedback physics \citep{trujillogomez:2019.minimum.cluster.mass}.

We find no evidence of a lower cut-off in the star cluster mass function in any of our models that is not simply consistent with simulation resolution -- the mass function exhibits power-law behaviour as far down as can be resolved (Figures \ref{fig:massfunc}, \ref{fig:normmassfunc}). Therefore, either no low-mass cutoff or shallowing exists in our solution, or if it does, it is at a mass that is insufficiently resolved. As such, our results concerning the low-mass cluster initial mass function are inconclusive. But even with infinite mass resolution, addressing this question properly in numerical simulations will require some treatment of the granularity of stars, for both realistic stellar feedback and stellar dynamics. The IMF sampling effect proposed in \citet{trujillogomez:2019.minimum.cluster.mass} is not captured by a feedback treatment adopting IMF-averaged feedback rates, and either requires some sampling scheme \citep[e.g.][]{sormani:2016.imf.sampling, su:2017.discreteness}, or individually-resolved stars. And the approximation of collisionless stellar dynamics is inapplicable when the dynamical time is comparable to the relaxation time (stellar mass scales $<100 M_\odot$), so collisional stellar dynamics could potentially affect the assembly process of low-mass clusters.



\subsection{Globular cluster formation}
A long-standing problem in star cluster formation is why star clusters in excess of $10^6 M_\odot$ formed in the early history of the Milky Way, now present as globular clusters \citep{harris:1996.mw.gcs}, but the mass scale of the most massive young star clusters in the present-day Milky Way is two orders of magnitude less \citep{portegies-zwart:2010.starcluster.review}.

There are two possible pictures for the behaviour of the galactic cluster IMF that might explain this. First, observed young star cluster mass functions often have a $\propto M^{-2}$ form with no well-constrained truncation mass \citep[e.g.][]{chandar:2017.cfe, mok:2019,adamo:2020.cluster.review}. Therefore, assuming that star cluster formation can be understood as a galaxy-wide statistical process in which this mass function is being sampled, one would expect to sample only a few very massive clusters throughout the galactic history. Thus it is not necessarily required for the underlying distribution to change over cosmic time. In \S\ref{section:montecarlo} we found that while holding our mass function fixed, simply varying the SFR was sufficient to drive variations in the maximum observed mass at a given time in accordance with the observed relation with galactic SFR \citep{bastian:2008.cfe}, so sampling effects {\it are} clearly important to consider when deriving inferences about the e.g. the maximum star cluster mass.

The other possibility is that the initial cluster mass function observed over a certain of time in the galaxy does have a physical truncation imposed by the ISM conditions, as discussed in \ref{sec:discussion:cmf}, and is not well-described by a pure power-law. This is found in M31 \citep{johnson:2017.m31.massfunction} and possibly in M51 \citep{messa:2018.m51, mok:2019}. To produce the most massive clusters, this truncation mass would then have to vary over the history of the galaxy, from large values at early times to smaller values today. According to our model, this picture is likely to be a better description of the galactic cluster formation history based on what is known about how galactic ISM conditions did indeed vary over time. Specifically, high-redshift galaxies had larger gas fractions (allowing more massive GMCs to form) \citep{tacconi:2020.cosmic.ism} and higher mid-plane ISM pressures \citep{cafg:sf.fb.reg.kslaw, gurvich:2020.ism.pressure}, leading to higher GMC surface densities and SFEs. 

Within this second picture, an important detail to be clarified is the role of galactic mergers in generating the conditions that are favourable for GC formation. This should ultimately imprint upon the age-mass-metallicity statistics of the GC population. In cosmological simulations that trace the formation and evolution of GCs with a sub-grid model (E-MOSAICS), \citet{kruijssen:2019.emosaics} found that mergers are subdominant for GC formation. However, in simulations that resolved {\it individual} cluster formation with a {\it resolved} ISM, \citep{li:2019.cfe} found that mergers greatly enhance the efficiency of cluster formation and thus play an important role. \citet{kim:2018.fire.gcs} and \citet{ma:2019.hiz.gc} simulated the formation of {\it dynamically-resolved} globular clusters, and both found that mergers do play a special role in generating the conditions necessary to form the most massive GC candidates. In these works, the high-pressure conditions present during mergers appear to allow the formation of self-gravitating gas clouds with high surface density, which in turn leads to high $\epsilon_{\rm int}$ and $f_{\rm bound}$. However, those simulations could not be run to redshift 0, so it is not clear how these clusters relate to the population of GCs that are presently observed. They instead focus upon $z>5$, ie. before the $z\sim 2$ peak of GC formation predicted by E-MOSAICS \citep{reinacampos:2019.emosaics}, so results pertaining to the importance of mergers may not generalize to GC formation as a whole.




\subsection{Comparison with \citealt{kruijssen:2012.cluster.formation.efficiency}}


In \citet{adamo:2015.m83.clusters}, it was found that the \citet{kruijssen:2012.cluster.formation.efficiency} (hereafter \citetalias{kruijssen:2012.cluster.formation.efficiency}) model was able to predict the observed $f_{\rm bound}$ with an accuracy that is comparable to the present work (Fig \ref{fig:m83gamma}). It was then combined with the \citet{kruijssen:2014.cluster.formation} formula to predict the maximum star cluster mass (here using our notation):
\begin{equation}
    M_{\rm cl,max} = \epsilon_{\rm int} f_{\rm bound} M_{\rm Toomre},
\end{equation}
where $M_{\rm Toomre}$ is the maximum gas mass that can collapse according to the Toomre instability, and $\epsilon_{\rm int}$ was given an assumed fiducial value of $5\%$. Again, $M_{\rm cl,max}$ was predicted with an accuracy comparable to the present work (cf. Figure \ref{fig:m83mmax}). It is illustrative to compare and contrast this framework with the one in the present work. \citetalias{kruijssen:2012.cluster.formation.efficiency} and this work present the same overall physical picture: hierarchical star formation produces stars over a wide range of densities, and in denser conditions feedback is less able to moderate star formation, leading to more efficient bound star cluster formation. However, the models do have important quantitative differences regarding the details of feedback-moderated star formation.

To summarize, \citetalias{kruijssen:2012.cluster.formation.efficiency} modeled the galactic ISM using the density statistics of isothermal, supersonic turbulence \citep{km2005}, which are fixed by three galactic bulk parameters: the Toomre stability parameter $\mathcal{Q}$, the mean disk gas surface density $\Sigma_{\rm gas}$, and the orbital frequency $\Omega$. These determine a log-normal gas density PDF which, according to the hierarchical star formation paradigm, maps onto the distribution of densities $\rho$ at which stars form. A feedback time-scale $t_{\rm fb}$ is introduced, identified with the time required for feedback to disrupt a gas overdensity and halt star formation, on the order of several $\mathrm{Myr}$, with only weak residual dependence on the three parameters (note that alternate feedback formulations can be slotted into the model, and this is merely the fiducial model). Locally-high $\epsilon_{\rm int}$ occurs in the upper tail of the gas density distribution where $t_{\rm ff} \ll t_{\rm fb}$, as star formation can proceed with a per-freefall efficiency of $\sim 1\%$ for many freefall times until $\epsilon_{\rm int} \sim 1$ locally, and the bound fraction can be correspondingly high.

This picture starts with the same premises (ISM physics and stellar feedback) and arrives at the same conclusion (locally-high SFE and bound cluster formation in dense regions) as the present work, but some differences should be noted between what is assumed in the analytic calculations, and what is found our simulations, and other recent, qualitatively-similar simulations \citep{geen:2017, kim:2018, li:2019.cfe}. We find that the per-freefall SFE is not universal \citep{grudic:2016.sfe}, nor in any detailed agreement with turbulence-regulated derivations of $\epsilon_{\rm ff}$ such as \citet{km2005}, the different assumptions considered in \citetalias{kruijssen:2012.cluster.formation.efficiency}. Simulations of feedback-moderated star formation on GMC scales typically find that $\epsilon_{\rm ff}$ and $\epsilon_{\rm int}$ are intimately linked to one another \citep[consistent with recent observations, see][]{kruijssen:2019.ngc300,chevance:2020.gmcs}, and scale with cloud parameters in a manner similar to Equation \ref{eq:sfefit}. Hence the value $\epsilon_{\rm ff} \sim  1\%$ is expected to be emergent, and to depend sensitively upon stellar feedback physics. The dimensional scalings of the {\it fiducial} feedback model in \citetalias{kruijssen:2012.cluster.formation.efficiency} and the present work also differ. The quantity $t_{\rm fb}$ is a characteristic {\it time-scale} that determines SFE and $f_{\rm bound}$, while the characteristic quantities $\Sigma_{\rm crit}$ and $\Sigma_{\rm bound}$ in our model predict scalings with GMC gas surface density. \citetalias{kruijssen:2012.cluster.formation.efficiency} did consider an alternate model with a similar scaling, formulating an assuming instantaneous radiative feedback without any time delay, but found that this predicted values of $f_{\rm bound}$ that would be difficult to distinguish from their fiducial model in observations. Therefore, while numerical simulations could potentially inform extensions or corrections to the assumptions made in \citetalias{kruijssen:2012.cluster.formation.efficiency}, it is not clear that its results would be strongly affected. This suggests that the elements in common between \citetalias{kruijssen:2012.cluster.formation.efficiency} and the present work represent the key physics driving stellar cluster formation.

\section{Summary and future work}\label{sec:conclusion}

In this work we have used numerical simulations of star-forming GMCs to explore the mapping between GMCs and the star clusters that they form. We have found that mapping is complex, and not one-to-one, due to of the variety of outcomes made possible by stochastic variations in the  internal turbulent flows of the clouds. In essence, the overall SFE of GMCs is reasonably predictable because the efficiency of feedback depends upon the \textit{macrostate} of the cloud, whereas cluster formation occurs on much smaller scales within the cloud, and is thus determined by the specific \textit{microstate} of the turbulent gas motions leading to star formation.

Despite this complexity, we have been able to explore the range of variations from one microstate to another and have found that the mapping from clouds to clusters does admit a statistical model (Section \ref{section:model}) that encodes fairly simple scalings in cluster formation efficiency, star cluster sizes, and star cluster masses. When we apply this model to a real population of GMCs, we successfully predict the fraction of star formation in bound clusters, the maximum cluster mass, and the size distribution of massive clusters. This is one of the first instances in which numerical simulations have succeed at reproducing the properties of star clusters in detail, using the observed GMC properties as initial conditions. Our key findings are as follows:
\begin{itemize}
\item Essentially all GMCs will form some fraction of their stars in bound clusters -- the formation of bound clusters and unbound associations are part of the same continuum, and not distinct processes \citep[e.g.][]{kruijssen:2012.cluster.formation.efficiency}. The star formation efficiency $\epsilon_{\rm int}$ and the bound fraction of star formation $f_{\rm bound}$ are correlated, but distinct quantities (Figure \ref{fig:SFEvCFE}). Both scale as an increasing function of the cloud surface density $\Sigma_{\rm GMC}$, eventually saturating to an order-unity value \citep[eg.][]{fall:2010.sf.eff.vs.surfacedensity,  murray:molcloud.disrupt.by.rad.pressure, grudic:2016.sfe}. 
\item $f_{\rm bound}$ scales steeply with $\Sigma_\mathrm{GMC}$, from a typical value of a few per cent in GMCs with surface density $\sim\unit[50]{\mspc}$ in our galaxy \citep{goddard:2010.cfe}, to $10-30\%$ in nearby spiral galaxies in which GMC surface densities are systematically a factor of $\sim 2$ higher \citep{faesi.lada:2018.gmcs}. The cloud-scale $f_{\rm bound}$ saturates to $\sim 1$ when the cloud scale SFE is $10-20\%$, in good agreement with \citet{li:2019.cfe}.
\item For a given set of cloud parameters, $f_{\rm bound}$ exhibits large variations, especially at low $\Sigma_{\rm GMC}$. We construct a statistical model that reproduces the scatter in simulation results shown in Figure \ref{fig:CFE} (Equations \ref{eq:sigmaprime} and \ref{eq:cfefit2}).
\item GMCs generally form multiple bound clusters, with masses distributed according to a cloud-scale mass distribution (Equation \ref{eq:massfuncfit}). The primary cluster tends to have a large (10-90\%) of the the total bound mass.
\item Stellar feedback is crucial in setting star cluster properties. Radiation is the most important, stellar winds are somewhat important, and SNe are essentially irrelevant because they come too late (see Table \ref{table:physicstests}). Because cluster masses are so sensitive to the strength of feedback (Figure \ref{fig:m83massfunc}), they provide a tight observational constraint on it.
\item The formation of bound clusters from $1\%$ solar metallicity gas is more efficient than at solar metallicity (Figure \ref{fig:CFE}). We have isolated this effect to the weaker stellar wind feedback expected from low-metallicity OB stars (Section \ref{sec:physicstests}), whose winds are effectively irrelevant compared to other feedback mechanisms.
\item The weak size-mass relation of star clusters is set during the star formation process, which produces large intrinsic scatter in cluster radii. There is a relation, however: star clusters from a given cloud form with a 3D density that depends upon the bulk properties of the parent cloud (Equation \ref{eq:sizefit2}). While our predicted size function is subject to some numerical caveats (\S \ref{sec:sizemass}), our model still gives predictions in good agreement with observed cluster sizes (\S \ref{section:M83}).
\end{itemize}


Though our success in reproducing star cluster bulk properties is encouraging, the problem of star cluster formation is hardly solved. Because our method of simulating star formation is approximate, we anticipate that comparisons with observed cluster properties that go beyond simple bulk properties will reveal interesting discrepancies. The extragalactic observations that we have compared with are likely the {\it easiest} constraints to satisfy. Meanwhile, the detailed cluster kinematics, and temporal and spatial age distributions that can be observed in the Milky Way and its satellites may well provide more powerful constraints on the star formation process.

At this point, it is likely that the most worthwhile gains in simulation realism can only be made by attacking the harder version of the problem: resolving the formation and motion of individual stars self-consistently, rather than assuming the IMF and using a simple stellar population formalism. Our simulations are reaching the scales where the granularity of stars can easily become important, and the details of how and when individual stars form can have major implications for stellar feedback, and hence the subsequent cloud evolution \citep{elephant}. Due to computational cost, this has never been done on the scale of massive GMCs that can actually sample the IMF, and hence the effects of feedback from massive stars have yet to be demonstrated in a fully self-consistent calculation that {\it predicts} massive star formation with conclusive numerical convergence. However, the advent of massively-scalable, Lagrangian codes with fast, accurate MHD methods and well-developed feedback coupling techniques will soon make this possible: in \citet{guszejnov:2020.isothermal.mhd} we simulated GMCs up to $2\times 10^6 M_\odot$ with individually-resolved collapsing stellar cores without feedback, at a mass resolution 200 times finer than the present work. In future work we will present results with a full accounting of both protostellar and stellar feedback (Grudi\'{c} et al. 2020, in prep., Guszejnov et al. 2020, in prep.).

Questions of particular importance for the next generation of GMC simulations include the behaviour of protostellar and main-sequence feedback acting in concert (ie. does regulating star formation on small scales ultimately affect cloud-scale behaviour?), what are the necessary and sufficient physics to satisfy constraints on the IMF turnover mass \citep[e.g.][]{bate:2009.imf,krumholz:2011.imf,federrath:2017.imf, guszejnov:2016.imf.feedback,guszejnov:2019.imf.variations}, and what deviations from universality might be expected in different environments. These are questions that can only be addressed by a comprehensive, individually-resolved approach to stellar feedback and dynamics.

\section*{Data availability}
The data supporting the plots within this article are available on reasonable request to the corresponding author. A public version of the GIZMO code is available at \url{http://www.tapir.caltech.edu/~phopkins/Site/GIZMO.html}.

\section*{Acknowledgements}
We thank Erik Rosolowsky and Mark Krumholz for useful discussions and for providing data and analysis code from \citet{freeman:2017.m83.gmcs} and \citet{krumholz:2018.star.cluster.review} respectively. We thank Norman Murray, Charles Lada, Anna Rosen, Hui Li, Mark Vogelsberger, Bruce Elmegreen, Angela Adamo, Eve Ostriker, Jeong-Gyu Kim, Marta Reina-Campos, and Sebastian Trujillo-Gomez for enlightening discussions that informed and motivated this work. Support for MYG was provided by a CIERA Postdoctoral Fellowship. Support for PFH was provided by an Alfred P. Sloan Research Fellowship, NSF Collaborative Research Grant \#1715847 and CAREER grant \#1455342, and NASA grants NNX15AT06G, JPL 1589742, 17-ATP17-0214. CAFG was supported by NSF through grants AST-1412836, AST-1517491, AST-1715216, and CAREER award AST-1652522, by NASA through grant NNX15AB22G, and by a Cottrell Scholar Award from the Research Corporation for Science Advancement. JMDK gratefully acknowledges funding from the German Research Foundation (DFG) in the form of an Emmy Noether Research Group (grant number KR4801/1-1) and the DFG Sachbeihilfe (grant number KR4801/2-1), as well as from the European Research Council (ERC) under the European Union's Horizon 2020 research and innovation programme via the ERC Starting Grant MUSTANG (grant agreement number 714907). MBK acknowledges support from NSF CAREER award AST-1752913, NSF grant AST-1910346, NASA grant NNX17AG29G, and HST-AR-15006, HST-AR-15809, HST-GO-15658, HST-GO-15901, and HST-GO-15902 from the Space Telescope Science Institute, which is operated by AURA, Inc., under NASA contract NAS5-26555. This research was undertaken, in part, thanks to funding from the Canada Research Chairs program. Numerical calculations were run on the Caltech compute cluster ``Wheeler,'' allocations from XSEDE TG-AST130039 and PRAC NSF.1713353 (awards OCI-0725070 and ACI-1238993) supported by the NSF, and NASA HEC SMD-16-7592. 

This research has made use of use of NASA's Astrophysics Data System, \texttt{ipython} \citep{ipython}, \texttt {numpy}, \texttt {scipy} \citep{scipy}, and \texttt {matplotlib} \citep{matplotlib}. 




\bibliographystyle{mnras}
\bibliography{master} 




\appendix
\section{Resolution Effects upon Cluster Masses and Sizes}
\label{section:appendix}
In \citet{grudic:2016.sfe} we showed that our simulation setup predicts GMC-scale SFEs that are insensitive to the effects of numerical resolution (as well as many other details that are secondary to feedback, such as star formation prescription, cooling details, and magnetic fields). However, the predictions of this work regarding masses and sizes of bound clusters must also be examined for numerical effects.

\begin{figure}
    \centering
    \includegraphics[width=\columnwidth]{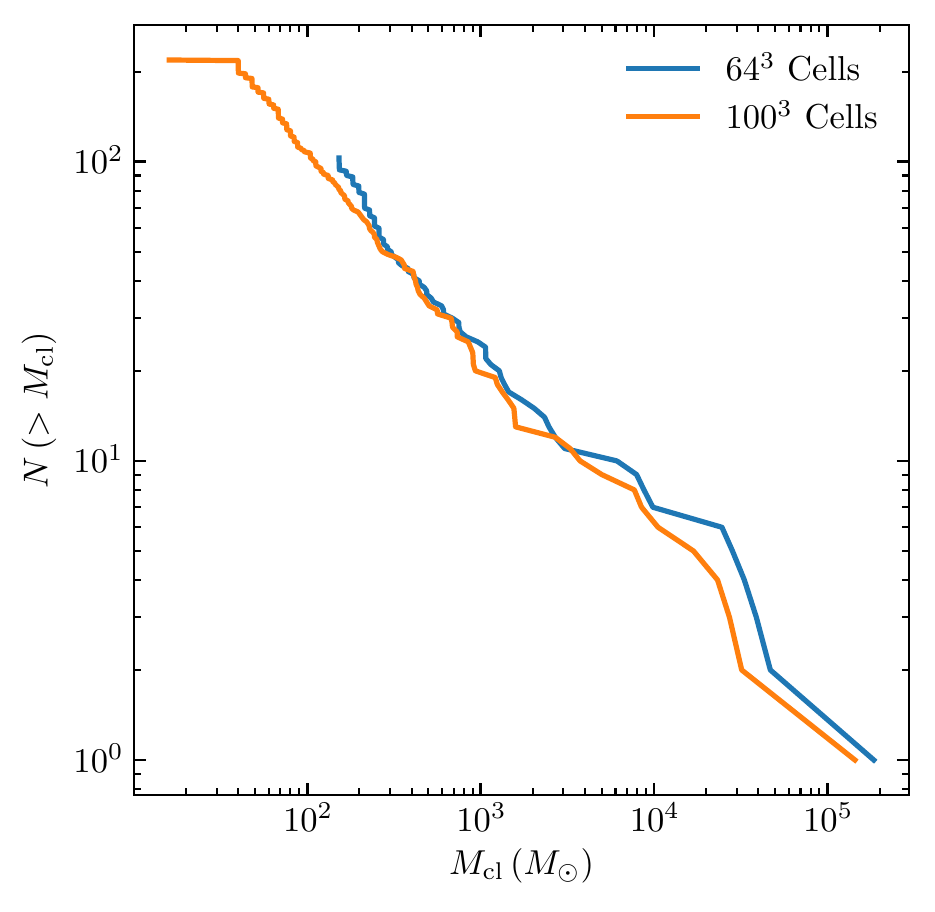}
    \caption{Cumulative mass distribution of the clusters formed in the 10 realizations of our GMC model with $M_\mathrm{GMC}=4\times 10^6 M_\odot$ , $R_\mathrm{GMC}=100\rm pc$, for two different mass resolutions: $100^3$ (ie. $\Delta m=4 M_\odot$), and $64^3$ (ie. $\Delta m=15.2M_\odot$).}
    \label{fig:massres}
\end{figure}

In Figure \ref{fig:massres} we plot the cluster mass functions stacked over our 10 statistical relizations of initial turbulence for our fiducial cloud model ($M_\mathrm{GMC}=4\times 10^6 M_\odot$ , $R_\mathrm{GMC}=100\rm pc$), at two different mass resolutions: $64^3$ Lagrangian gas cells, and $100^3$ cells (our fiducial resolution). We find that the mass functions agree well over their common resolved range, and the main difference is that the finer-resolved simulations extend to smaller masses. Hence our results for star cluster masses and $f_\mathrm{bound}$ are fairly insensitive to mass resolution except in the case where the cluster mass in question simply cannot be resolved at all.

\begin{figure}
    \centering
    \includegraphics[width=\columnwidth]{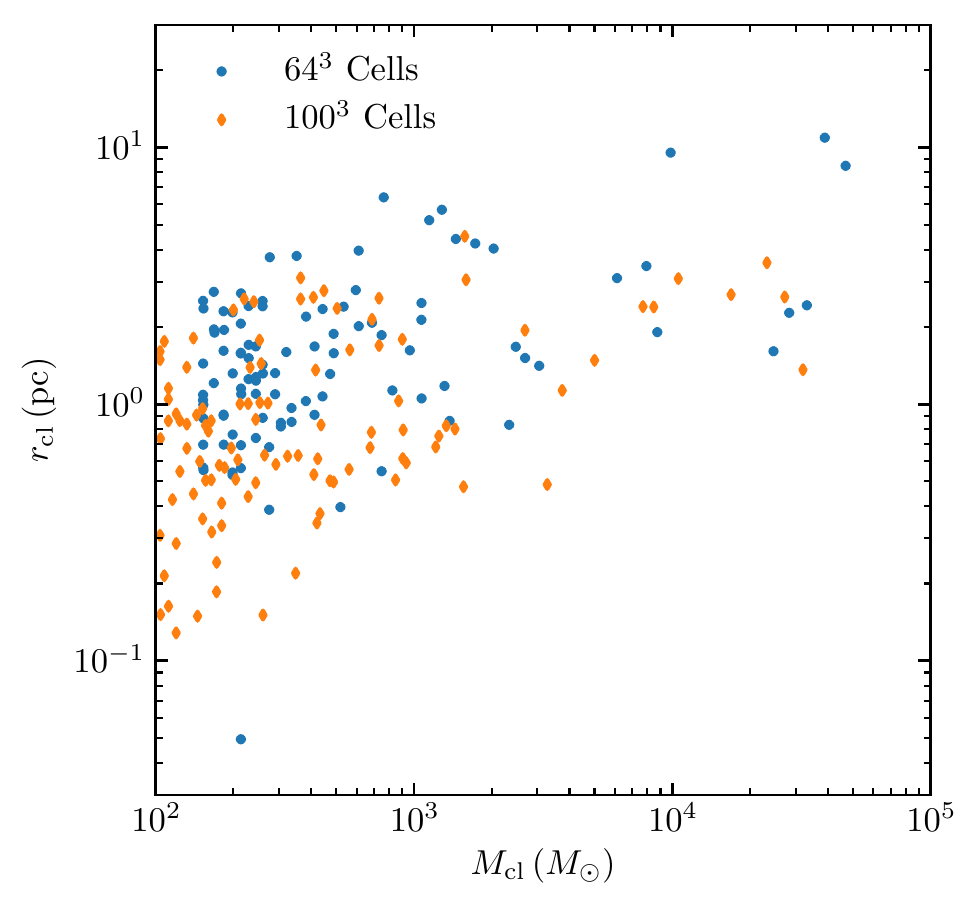}
    \caption{Size-mass relation of the clusters formed in the 10 realizations of our GMC model with $M_\mathrm{GMC}=4\times 10^6 M_\odot$ , $R_\mathrm{GMC}=100\rm pc$ for two different mass resolutions. Our numerical technique produces preferentially smaller clusters at higher resolution, possibly hinting at missing physics (see discussion in \ref{section:appendix}).}
    \label{fig:sizeres}
\end{figure}

In Figure \ref{fig:sizeres} we plot the size-mass relations for these same simulations at two different resolution levels, and find that clusters formed in finer-resolved simulations are preferentially more compact. Therefore, we cannot rule out that our cluster sizes are sensitive to numerical resolution, and hence that the cluster size predictions are not fully self-consistent. Indeed, it is not clear whether it is even {\it possible} to self-consistently predict star cluster sizes using the collisionless approximation that we have adopted. A collisionless self-gravitating fluid conserves its maximum phase-space density \citep{grudic:2017}, and this approximation breaks down at the scale where the collisionless approximation does, ie. in clusters where the relaxation time is comparable to the dynamical time. Therefore, we acknowledge that fully self-consistent predictions of star cluster sizes may well require a full N-body technique.



\bsp	
\label{lastpage}
\end{document}